\pdfoutput=1

\documentclass[11pt,a4paper]{article}

\usepackage{jheppub}
\usepackage{amsmath}
\usepackage{multicol,bbm}
\usepackage{enumerate}
\usepackage{bibentry}

\usepackage{CJKutf8}

\usepackage{graphicx}
\usepackage{dcolumn}
\usepackage{float}
\usepackage[many]{tcolorbox}
\usepackage{lipsum}
\usepackage{cancel}

\usepackage{shuffle}
\usepackage{graphbox}
\usepackage{environ}
\usepackage{physics}

\usepackage{amssymb}
\usepackage{amsthm}

\usepackage{bm}
\usepackage{multirow}
\usepackage{caption}
\usepackage{tabularx}
\usepackage{enumitem}

\usepackage{extarrows}
\usepackage{tikz}
\usepackage{verbatim} 

\usetikzlibrary{decorations.pathmorphing}
\usetikzlibrary{decorations.pathreplacing}

\graphicspath{{figures/}}

\begin{document}


\title{Two-loop MHV Form Factors from the Periodic Wilson Loop}


\date{\today}

\author[a]{Zhenjie Li (\begin{CJK*}{UTF8}{gkai}李振杰\end{CJK*})} 

\affiliation[a]{SLAC National Accelerator Laboratory, Stanford University, Stanford, CA 94309, USA}
\emailAdd{zhenjiel@slac.stanford.edu}


\date{\today}

\abstract{
We discuss how to compute maximal-helicity-violating (MHV) form factors for the chiral part of the stress-tensor supermultiplet from periodic light-like polygon Wilson loops in planar $\mathcal N=4$ super Yang-Mills theory beyond the one-loop level. We show that the periodicity imposes path ordering on points on different edges, which explains the appearance of square roots coming from non-planar Feynman diagrams. Taking such diagrams into account, we provide the integrand of the two-loop $n$-particle MHV form factor, compute all diagrams, prove the cancellation of divergences and finally compute the two-loop 5-particle and 6-particle form factors as examples.
}

\maketitle

\section{Introduction}

In the last decade, the duality between amplitudes and light-like polygon Wilson loops in planar ${\cal N} = 4$ supersymmetric Yang-Mills theory (sYM) has been shown to be a powerful duality at both weak and strong coupling~\cite{Alday:2007hr,Alday:2007he,Drummond:2007cf,Drummond:2007aua,Drummond:2007au,Brandhuber:2007yx,Drummond:2008aq,Bern:2008ap,DelDuca:2009au,DelDuca:2010zg,DelDuca:2010zp,CaronHuot:2010ek,Alday:2010ku,Mason:2010yk}. It not only provides a new understanding of some aspects of amplitudes, but also helps to compute amplitudes for higher multiplicities at weak coupling, \textit{e.g.} two-loop MHV and NMHV amplitude for any multiplicity~\cite{Anastasiou:2009kna,Caron-Huot:2011zgw, He:2020vob}. After its success for amplitudes, the duality has been further generalized to other gauge-invariant but off-shell quantities, \textit{e.g.} form factors \cite{Maldacena:2010kp,Gao:2013dza,Sever:2020jjx,Sever:2021nsq,Sever:2021xga,Brandhuber:2010ad,Basso:2023bwv} and correlation functions \cite{Alday:2010zy,Alday:2011ga,Eden:2010ce,Adamo:2011dq}.

\begin{figure}[t]
\centering
\begin{tikzpicture}[scale=0.5]
\draw[thick] (-3.5,-1) -- (-2.5,1) -- (-0.5,2) -- (1.5,1.5) -- (2.5,0.5) -- (3,-1);
\draw[thick] (3,-1) -- (4,1) -- (6,2) -- (8,1.5) -- (9,0.5) -- (9.5,-1);
\draw[thick] (9.5,-1) -- (10.5,1);
\draw[thick] (-3.5,-1) -- (-4,0.5);
\node at (-5,0) {$\cdots$};
\node at (11.5,0) {$\cdots$};
\draw (3.3,-0.1) -- (3.57,0.14) -- (3.54,-0.21);
\draw (2.01,1.17) -- (2.08,0.91) -- (1.82,0.97);
\draw (6.88,1.92) -- (7.1,1.72) -- (6.8,1.66);
\draw (-1.85,1.46) -- (-1.53,1.49) -- (-1.72,1.24);
\draw (9.3,0) -- (9.24,-0.20) -- (9.06,-0.06);
\node at (-3.5,-1.5) {$1$};
\node at (-2.8,1.4) {$2$};
\node at (3,-1.5) {$1^+$};
\node at (2.8,0.7) {$n$};
\node at (4,1.4) {$2^+$};
\node at (9.5,0.8) {$n^+$};
\end{tikzpicture}
\caption{The $n$-particle periodic Wilson loop. We fix the direction of the contour to be clockwise and omit it if there is no conflict. We also define that $i^\pm:=i\pm n$, then $x_{i^+}-x_i=q$ for any $i$.}
\label{fig:pWL}
\end{figure}
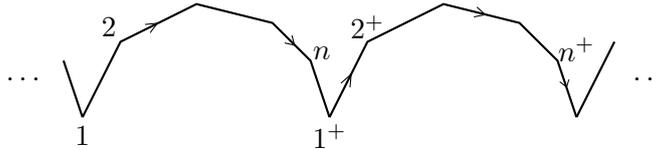

In this note, we focus on the duality associated to the MHV form factor in planar ${\cal N} = 4$ sYM
\begin{equation}
F\left(p_1, \ldots, p_n ; q\right):=\int d^4 x \,e^{-i q \cdot x}\left\langle p_1, \ldots, p_n\right| \mathcal{O}(x)|0\rangle
\end{equation}
of the lowest protected operator $\mathcal O(x)\sim\operatorname{Tr}(\phi^2)$ with the momentum conservation $q=p_1+\cdots+p_n$, 
where $\phi$ is a complex scalar field such that the operator falls in the same $\frac12$-BPS supermultiplet as the stress-energy tensor. It is first argued from the string picture~\cite{Maldacena:2010kp} that should be dual to a periodic Wilson loop (see Fig.~\ref{fig:pWL}),
\begin{equation}
\mathcal F:=F/ F^{\text{tree}} \sim \langle W[\mathcal{C}_n]\rangle=\langle \operatorname{Tr} \mathcal{P} \exp \left[i g_{\text{YM}} \oint_{\mathcal{C}_n} d x^\mu A_{\mu}^a(x)T^a\right]\rangle,
\end{equation}
where $\mathcal{P}$ is the path-ordering and the contour $\mathcal C_n=\bigcup_{i=-\infty}^\infty \ell_i$ is the infinite union of light-like edges
\begin{equation}
\ell_i:=\left\{x^\mu\left(t_i\right)=(1-t_i) x_i^\mu+t_i x_{i+1}^\mu \,:\, t_i \in[0,1]\right\},
\end{equation}
where $x_{i+1}-x_i=p_i$ is light-like with the requirement that $x_{i+n}-x_i = q$ for every $i$ to make the Wilson loop periodic. We further need to take the large-$N_c$ limit and expand it in the ’t Hooft coupling $g^2=g_{\text{YM}}^2N_c/(4\pi)^2$ for the gauge group $\operatorname{SU}(N_c)$, and then its result is a function of planar variables 
\begin{equation}
    x_{ij}^2:=(x_i-x_j)^2=(p_i+\cdots+p_{j-1})^2=s_{i,\dots,j-1}.
\end{equation}
In the string picture, the periodicity of the Wilson loop comes from the non-trivial topology caused by the
insertion of a closed string state on the worldsheet, which requires to extend the space to its universal cover periodically after T-duality.

This duality was first developed and checked at strong coupling \cite{Maldacena:2010kp,Gao:2013dza}, and then an OPE for the form factors \cite{Sever:2020jjx,Sever:2021nsq,Sever:2021xga} based on this duality was developed at finite coupling, and matches results at both strong and weak coupling. In \cite{Basso:2023bwv}, the duality has been generalized to form factors of all $\frac12$-BPS operators. 
However, at weak coupling, this duality has been only checked at tree~\cite{Koster:2016loo,Koster:2016fna} and one-loop level~\cite{Brandhuber:2010ad,Brandhuber:2011tv}.


Recently, the perturbative expansion of the MHV form factors for $\operatorname{Tr}(\phi^2)$ at weak coupling has been pushed to a high level. The three-particle MHV form factors have been bootstrapped to eight loops~\cite{Dixon:2022rse}, and the two-loop four-particle MHV form factor was computed through the bootstrap~\cite{Dixon:2022xqh,Dixon:2024yvq} and Feynman integrals \cite{Guo:2024bsd}. Meanwhile, from these results, a novel and unexpected duality in parity-even kinematics, antipodal duality, has been first discovered between three-particle form factors and six-particle MHV amplitudes~\cite{Dixon:2021tdw}, and then this duality was found to be a consequence of the antipodal \textit{self}-duality of the four-particle form factor~\cite{Dixon:2022xqh}.

In this note, we begin the investigation of the form factor/periodic Wilson loop duality beyond one loop at weak coupling, which not only checks the duality, but also provides some data points to help understanding antipodal duality in the future.

Since the weak-coupling expansion of a light-like closed polygon bosonic Wilson loop at multiloop level has been well studied in \cite{Drummond:2007au,Drummond:2008aq,Anastasiou:2009kna,DelDuca:2009au,DelDuca:2010zg,DelDuca:2010zp}, it might seem that its generalization to a periodic one is straightforward. In fact, this is the way to get the one-loop form factor in \cite{Brandhuber:2010ad}. The one-loop expansion of Wilson loop only has one type of diagram
\begin{equation}
I_{\text{O}}(i,j)=
\begin{tikzpicture}[baseline={([yshift=-.5ex]current bounding box.center)},scale=0.55]
\draw[thick] (-2.5,2.5) -- (-3,-0.5);
\draw[thick] (0.5,2.5) -- (1,-0.5);
\draw[decorate,decoration={aspect=0.3,coil,segment length=4,amplitude=3}] 
        (0.6887,1.4367) .. controls (-0.5,0.5) and (-1.5,0.5) .. (-2.6806,1.5036);
\node at (-3.4,-0.8) {$x_i$};
\node at (0.9,2.8) {$x_j$};
\node at (-3.0,2.8) {$x_{i+1}$};
\node at (1.4,-0.9) {$x_{j+1}$};
\node[circle,inner sep=1pt,fill=black] at (1,-0.5) {};
\node[circle,inner sep=1pt,fill=black] at (-3,-0.5) {};
\node[circle,inner sep=1pt,fill=black] at (0.5,2.5) {};
\node[circle,inner sep=1pt,fill=black] at (-2.5,2.5) {};
\node at (-3,1.5) {$z_i$};
\node at (1,1.5) {$z_j$};
\end{tikzpicture}
=\int_{x_i}^{x_{i+1}}\!\!\int_{x_j}^{x_{j+1}} dz_i^\mu dz_j^\nu D^{\mu\nu}(x-y),
\end{equation}
where 
\begin{equation}
D^{\mu\nu}(x-y):=\langle A^\mu(x) A^\nu(y)\rangle_0=-\frac{\pi^{\epsilon}}{4 \pi^2} \frac{\Gamma(1-\epsilon)}{\left(-(x-y)^2+i 0^+\right)^{1-\epsilon}}\eta^{\mu\nu}
\end{equation}
is the coordinate space propagator in Feynman gauge. For one-loop amplitudes, one only needs to take the sum over all pairs of edges $\mathcal{A}^{(1)}_n=\sum_{1\leq i<j\leq n}I_{\text{O}}(i,j)$. For the one-loop form factor, the only difference is that we do not consider all infinite pairs of edges but only consider pairs of edges that can be fit into one period, then 
\begin{equation}\label{Foneloop}
\mathcal{F}^{(1)}_n=\sum_{i=1}^n \sum_{j=i+1}^{i+n-1}I_{\text{O}}(i,j).
\end{equation}
Therefore, a naive conjecture is that we use similar two-loop diagrams for closed Wilson loop/MHV amplitudes listed in \cite{Anastasiou:2009kna}, but only sum the diagrams can be contained in one period. However, we will face two main problems.

The first problem is the missing of a new type of square root. From the known analytic result of the two-loop four-particle MHV form factor~\cite{Dixon:2022xqh, Dixon:2024yvq,Guo:2024bsd}, we know that there are two types of square roots. The familiar one comes from one-loop triangle Feynman diagrams corresponding to the three-gluon vertex in the expansion of Wilson loops, which is already considered in the two-loop expansion of closed Wilson loops:
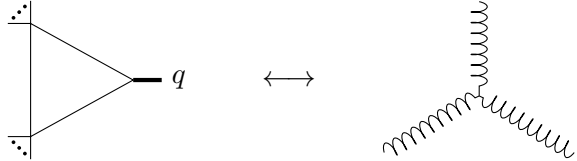
\[
\begin{tikzpicture}[baseline={([yshift=-.5ex]current bounding box.center)},scale=0.75]
\draw[ultra thick] (2.8,0.2) -- (3.3,0.2);
\node[inner sep= .5pt, fill = black, circle] at (0.7,1.3) {};
\node[inner sep= .5pt, fill = black, circle] at (0.7,-0.9) {};
\node[inner sep= .5pt, fill = black, circle] at (0.8,-1) {};
\node[inner sep= .5pt, fill = black, circle] at (0.9,-1.1) {};
\node[inner sep= .5pt, fill = black, circle] at (0.9,1.5) {};
\node[inner sep= .5pt, fill = black, circle] at (0.8,1.4) {};
\draw (1,1.2) -- (1,-0.8) -- (2.8,0.2) -- cycle;
\draw (0.6,-0.8) -- (1,-0.8) -- (1,-1.2);
\draw (0.6,1.2) -- (1,1.2) -- (1,1.6);
\node at (3.6,0.2) {$q$};
\end{tikzpicture}\qquad
\longleftrightarrow
\qquad
\begin{tikzpicture}[baseline={([yshift=-0.5ex]current bounding box.center)},scale=0.5]
\draw[decorate,decoration={aspect=0.3,coil,segment length=4,amplitude=3}] 
        (-2,-2) .. controls (-1,-1.5) and (-0.5,-1) .. (0.5,-0.5);
\draw[decorate,decoration={aspect=0.3,coil,segment length=4,amplitude=3}] 
        (3,-2) .. controls (2,-1.5) and (1.5,-1) .. (0.5,-0.5);
\draw[decorate,decoration={aspect=0.3,coil,segment length=4,amplitude=3}] 
        (0.5,2) .. controls (0.5,1) and (0.5,0.5) .. (0.5,-0.5);
\end{tikzpicture}
\]
However, there is a new square root $\Sigma_5\propto\sqrt{(1-u-v)^2-4 u v}$ for 
\[
u=\frac{s_{34}s_{41}}{s_{12}s_{23}},\quad v=\frac{4(q\cdot p_1)(q\cdot p_3)}{s_{12}s_{23}}
\]
in the two-loop four-particle MHV form factor.
This square root (and its cyclic image) comes from non-planar Feynman diagrams with 4 massless legs and 1 massive leg \cite{Abreu:2023rco,Bourjaily:2019gqu} like
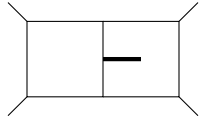
\begin{center}
\begin{tikzpicture}[scale=0.5]
\draw (-3.5,2) -- (-3.5,0) -- (-1.5,0) -- (-1.5,2) -- cycle;
\draw (-1.5,0) -- (0.5,0) -- (0.5,2) -- (-1.5,2);
\draw[ultra thick] (-1.5,1) -- (-0.5,1);
\draw (-4,2.5) -- (-3.5,2);
\draw (1,-0.5) -- (0.5,0);
\draw (-4,-0.5) -- (-3.5,0);
\draw (1,2.5) -- (0.5,2);
\end{tikzpicture}
\end{center}
but it is not predicted from any known two-loop Wilson loop diagram. 

The second problem is more crucial and difficult to detect: even the one-loop expansion eq.\eqref{Foneloop} is naively \textit{not} gauge invariant!
To see it, consider the massless Feynman propagator in the $R_\xi$-gauge
\begin{equation}\label{Dxi}
D_{\xi}^{\mu\nu}(x-y):=\langle A^\mu (x)A^\nu(y)\rangle=\int \frac{d^Dk}{k^2+i 0^+}\left(\eta^{\mu \nu}-(1-\xi) \frac{k^\mu k^\nu}{k^2}\right)e^{-ik\cdot (x-y)},
\end{equation}
and the $(1-\xi)$ part gives an extra contribution to $I_{\text{O}}(i,j)$ in the dimensional regularization
\begin{equation}\label{Sxi}
\begin{aligned}
S_\xi(i,j):&=\frac{\pi^{\epsilon}\Gamma(1-\epsilon)}{4 \pi^2} \frac{1-\xi}{4\epsilon}\bigl[(-x_{i,j}^2)^{\epsilon}+(-x_{i+1,j+1}^2)^{\epsilon}-(-x_{i+1,j}^2)^{\epsilon}-(-x_{i,j+1}^2)^{\epsilon}\bigr],
\end{aligned}
\end{equation}
where $x_{i,j}:=x_i-x_j$. This contribution should cancel for a gauge invariant object. To see it clearly, we denote a linear combination of four terms in $S_\xi(i,j)$ by a box
\[
\begin{tikzpicture}[scale=1,baseline={([yshift=-0.5ex]current bounding box.center)}]
\draw (0.5,-2) -- (0.5,-3) -- (1.5,-3);
\draw (1.5,-3) -- (1.5,-2) -- (0.5,-2);
\node at (0.65,-2.85) {\tiny $s_1$};
\node at (1.35,-2.85) {\tiny $s_2$};
\node at (1.35,-2.15) {\tiny $s_3$};
\node at (0.65,-2.15) {\tiny $s_4$};
\node at (1,-2.5) {$ij$};
\end{tikzpicture} = s_1(-x_{i,j}^2)^{\epsilon}+s_2(-x_{i+1,j}^2)^{\epsilon}+s_3(-x_{i+1,j+1}^2)^{\epsilon}+s_4(-x_{i,j+1}^2)^{\epsilon},
\]
where $s_a=-1,0,1$ for $a=1,2,3,4$ is the sign, and we set $s_a=0$ if $x_{m,n}^2=0$ at that corner. For example, for the closed polygon light-like Wilson loop with $4$ edges which is dual to the $4$-particle MHV amplitude, $\sum_{1\leq i<j\leq 4} S_\xi(i,j)$ is propotional to the diagram
\[
\begin{tikzpicture}[scale=1,baseline={([yshift=-0.5ex]current bounding box.center)}]
\draw (0.5,0) -- (0.5,-3) -- (1.5,-3);
\draw (0.5,0) -- (3.5,0) -- (3.5,-1);
\draw (0.5,-1) -- (2.5,-1) -- (3.5,-1);
\draw (0.5,-2) -- (2.5,-2) -- (2.5,0);
\node at (1,-2.5) {12};
\node at (1,-1.5) {13};
\node at (1,-0.5) {14};
\node at (2,-1.5) {23};
\node at (2,-0.5) {24};
\node at (3,-0.5) {34};
\node at (0.6,-2.9) {\tiny $0$};
\node at (2.4,-1.1) {\tiny $0$};
\node at (1.6,-1.9) {\tiny $0$};
\node at (1.4,-1.1) {\tiny $+$};
\node at (0.6,-1.9) {\tiny $+$};
\node at (1.4,-2.1) {\tiny $0$};
\node at (2.6,-0.9) {\tiny $0$};
\node at (3.4,-0.1) {\tiny $0$};
\node at (0.6,-0.9) {\tiny $+$};
\node at (1.4,-0.1) {\tiny $+$};
\node at (1.6,-0.9) {\tiny $+$};
\node at (2.4,-0.1) {\tiny $+$};
\node at (2.4,-0.9) {\tiny $0$};
\node at (0.6,-1.1) {\tiny $-$};
\node at (1.4,-1.9) {\tiny $0$};
\node at (0.6,-0.1) {\tiny $-$};
\node at (1.4,-0.9) {\tiny $-$};
\node at (0.6,-2.1) {\tiny $-$};
\node at (1.4,-2.9) {\tiny $0$};
\node at (1.6,-1.1) {\tiny $-$};
\node at (2.4,-1.9) {\tiny $0$};
\node at (3.4,-0.9) {\tiny $0$};
\node at (2.6,-0.1) {\tiny $-$};
\node at (1.6,-0.1) {\tiny $-$};
\draw (1.5,0) -- (1.5,-3);
\end{tikzpicture} = -(-x_{1,5}^2)^\epsilon = -(-x_{1,1}^2)^\epsilon = 0,
\]
where we glue boxes by placing corners with the same planar variable $x_{m,n}^2$ together, 
and we can see clearly that only the upper left corner of this diagram does not vanish naively, but $x_{1,5}=x_{1,1}=0$ because the Wilson loop is closed. However, for the three-particle form factor expansion in eq.\eqref{Foneloop}, it gives the non-vanishing gauge dependent part proportional to
\begin{equation}
\begin{tikzpicture}[scale=1,baseline={([yshift=-0.5ex]current bounding box.center)}]
\draw (0.5,-1) -- (0.5,-3) -- (1.5,-3);
\draw (1.5,0) -- (3.5,0) -- (3.5,-1);
\draw (0.5,-1) -- (2.5,-1) -- (3.5,-1);
\draw (0.5,-2) -- (2.5,-2) -- (2.5,0);
\node at (1,-2.5) {12};
\node at (1,-1.5) {13};
\node at (3,0.5) {$32^+$};
\node at (2,-1.5) {23};
\node at (2,-0.5) {$21^+$};
\node at (3,-0.5) {$31^+$};
\node at (0.6,-2.9) {\tiny $0$};
\node at (2.4,-1.1) {\tiny $0$};
\node at (1.6,-1.9) {\tiny $0$};
\node at (1.4,-1.1) {\tiny $+$};
\node at (0.6,-1.9) {\tiny $+$};
\node at (1.4,-2.1) {\tiny $0$};
\node at (2.6,-0.9) {\tiny $0$};
\node at (3.4,-0.1) {\tiny $0$};
\node at (2.6,0.1) {\tiny $+$};
\node at (3.4,0.9) {\tiny $+$};
\node at (1.6,-0.9) {\tiny $+$};
\node at (2.4,-0.1) {\tiny $+$};
\node at (2.4,-0.9) {\tiny $0$};
\node at (0.6,-1.1) {\tiny $-$};
\node at (1.4,-1.9) {\tiny $0$};
\node at (2.6,0.9) {\tiny $-$};
\node at (3.4,0.1) {\tiny $0$};
\node at (0.6,-2.1) {\tiny $-$};
\node at (1.4,-2.9) {\tiny $0$};
\node at (1.6,-1.1) {\tiny $-$};
\node at (2.4,-1.9) {\tiny $0$};
\node at (3.4,-0.9) {\tiny $0$};
\node at (2.6,-0.1) {\tiny $-$};
\node at (1.6,-0.1) {\tiny $-$};
\draw (1.5,0) -- (1.5,-3);
\draw (2.5,0) -- (2.5,1) -- (3.5,1) -- (3.5,0);
\node at (7.3,-0.1) {$-(-x_{1,1^+}^2)^\epsilon-(-x_{2,2^+}^2)^\epsilon-(-x_{3,3^+}^2)^\epsilon$};
\node at (7.3,-0.8) {$(-x_{2,1^+}^2)^\epsilon+(-x_{3,2^+}^2)^\epsilon+(-x_{1^+,3^+}^2)^\epsilon$};
\node at (4,-0.5) {$=$};
\node at (7.2,-1.6) {$\rotatebox{90}{=}$};
\node at (7.6,-2.4) {$\displaystyle\log\biggl(\frac{s_{12}s_{23}s_{31}}{(q^2)^3}\biggr)+O(\epsilon^1)$};
\end{tikzpicture}
\end{equation}
where $i^+:=i+n=i+3$ and $x_{i,i^+}^2=q^2$. Form factors are of course gauge invariant, so it only means that the sum of these diagrams is not exactly the integrand. 

In this note, we intend to solve these two problems and show how to compute two-loop MHV form factors via a periodic Wilson loop. In Section 2, we show how to solve these two problems from a correct understanding of the periodicity of the Wilson loop.  In Section 3, we review the definition of remainder functions and discuss which two-loop Wilson loop diagrams we need to compute by duality. In Section 4, we compute these diagrams. In Section 5, we compute the two-loop form factors of 5-particles and 6-particles as examples and discuss some properties of remainder functions for any multiplicity.

\paragraph{Ancillary files} We provide ancillary files to collect some results in this note. They are all in the Wolfram language and can be read by Mathematica directly. The file {\tt FF52L.m} ({\tt FF62L.m}) is the symbol of the remainder function of two-loop 5-particle (6-particle) MHV form factor, and the definition of their alphabets can be found in {\tt FF52L\_alphabet.m} and {\tt FF62L\_alphabet.m}. The 2D limit of {\tt FF62L.m} is given in the file {\tt FF62L\_2D.m}. We also provide explicit results of diagrams defined in Section \ref{sec:diags}. Their file names look like {\tt {\it name}\_diag\_{\it points}.m}, where {\it points} are used to label their edges. For example, the file {\tt star\_diag\_258.m} gives the result of the star diagram (defined in Section \ref{sec:star}) with three edges $[x_2,x_3]$, $[x_5,x_6]$ and $[x_8,x_9]$. One can get the result of a generic star diagram from it by replacing edges. For divergent diagrams, we provide their regulated version. 

\section{What can we learn from periodicity?}

In this section, we explain how to understand the periodicity of the Wilson loop. We first explain the origin of periodicity from the string picture by T-duality, and then discuss the consequence of periodicity and why it solves both problems stated in the Introduction.

\subsection{String picture and T-duality}\label{Tdual}

\begin{figure}
\centering
\begin{tikzpicture}[scale=0.75]
\draw[thick] (-6.5,1) -- (-6.5,-4);
\draw[color=gray!10!white, fill=gray!10!white] (1,0.5) -- (-1,1.5) -- (-1,-3.5) -- (1,-4.5) -- cycle;
\draw  (-6.5,-1) ellipse (0.1 and 0.3);
\draw  plot[smooth, tension=.7] coordinates {(-0.5,-0.5) (0,-1) (0.5,0.5)};
\draw  plot[smooth, tension=.7] coordinates {(-0.6,-0.8) (-0.2,-2) (-0.6,-3)};
\draw  plot[smooth, tension=.7] coordinates {(0.6,0.2) (0.2,-1.8) (0.8,-2.4)};
\draw  plot[smooth, tension=.7] coordinates {(0.8,-2.6) (0,-2.5) (-0.6,-3.4)};
\draw  plot[smooth, tension=.7] coordinates {(-6.5,-0.7) (-4,-0.3) (-1.6,-0.9) (-0.4,-1.5) (0,-1.2)};
\draw  plot[smooth, tension=.7] coordinates {(-6.5,-1.3) (-3.9,-2.1) (-1.4,-2) (-0.4,-2.1) (-0.1,-2.4)};
\node at (-6.5,-4.5) {boundary $z=0$};
\node at (0.5,-4.5) {$z=z_{\text{IR}}>\!\!\!>0$ near horizen};
\end{tikzpicture}
\caption{The string picture. We have a closed string at the boundary and open strings ending at a D3 brane near the horizon of AdS$_{5}$.}
\label{fig:string}
\end{figure}
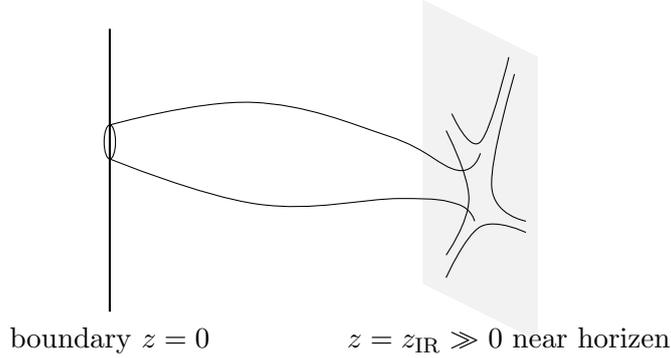

To understand the origin of periodicity, we go back to its string picture, see Fig.~\ref{fig:string}, and apply T-duality to get the Wilson loop. 

Following the T-duality 
procedure for MHV amplitudes (closed Wilson loops) in \cite{McGreevy:2007kt} and for non-planar amplitudes \cite{Ben-Israel:2018ckc}, 
we consider the AdS$_5$ background $ds^2=(dz^2+dx^2)/z^2$. There are $n$ vertex operators $V_i\sim \exp(-ik_i\cdot x(0,\sigma_i))$ located at $\tau=0$ for open string states and a closed string state $V_{0}\sim \exp(iq\cdot x(-\infty,\sigma_{0}))$ located at $\tau=-\infty$, so the Euclidean worldsheet action looks like 
\begin{equation}
S=\frac{\sqrt{\lambda}}{4 \pi} \int_{\mathcal{D}} d \sigma d \tau \frac{(\partial_\alpha z)^2+(\partial_\alpha x)^2}{z^2}-i \sum_{i=1}^n k_i \cdot x(0, \sigma_i)+i  q \cdot x(-\infty,\sigma_{0}),
\end{equation}
where $\alpha$ is the worldsheet index. 
By introducing the dual momentum $K_i=\sum_{j=1}^ik_j$, the second term can be rewritten as 
\[
\begin{aligned}
\sum_{i=1}^n k_i \cdot x(0, \sigma_i)&=q\cdot x(0,\sigma_{n+1})+\sum_{i=1}^{n}K_{i} \cdot (x(0, \sigma_i)-x(0, \sigma_{i+1}))\\
&=q\cdot x(0,\sigma_{n+1})-\sum_{i=1}^{n}K_{i} \cdot \int^{\sigma_{i+1}}_{\sigma_i} d\sigma\,\partial_\sigma x(0,\sigma),
\end{aligned}
\]
where we introduce $\sigma_{n+1}=\sigma_1+2\pi-0^+$ to make the integral almost go around the circle, 
and similarly we can rewrite the difference $x(0,\sigma_{n+1})-x(-\infty,\sigma_{0})$ as a integral of $\partial_\tau x$ by introducing a path $\gamma:[-\infty,0]\mapsto [\sigma_0,\sigma_{n+1}]$ connecting $\sigma_0$ with $\sigma_{n+1}$, then the action becomes
\[
S=\frac{\sqrt{\lambda}}{4 \pi} \int_{\mathcal{D}} d \sigma d \tau \frac{(\partial_\alpha z)^2+(\partial_\alpha x)^2}{z^2}+i \sum_{i=1}^n K_i \cdot \int^{\sigma_{i+1}}_{\sigma_i} d\sigma\,\partial_\sigma x(0,\sigma)-i  q \cdot \int_{-\infty}^0 \!\!\!d\tau \,\partial_\tau x(\tau,\gamma(\tau)).
\]

To perform the T-duality, we first notice that the action $S$ is invariant under the transition $x\to x+a$ for a constant $a$, so we can introduce a minimal coupling to a gauge field $A_\alpha$ transforming as $A_\alpha\to A_\alpha+\partial_\alpha a$ when $x\to x+a$, and then replace all $\partial_\alpha x$ with $\partial_\alpha x-A_\alpha$ in the action. Besides, to avoid nontrivial topological effects and the dependence on the choice of $\gamma$, we also need to impose the condition that holonomy $\oint d\sigma A_\sigma(\tau,\sigma)$ vanishes for all $\tau$, it can be furnished by adding the following terms
\[
i\int_{\mathcal{D}} d \sigma d \tau\, y\cdot (\partial_\tau A_\sigma-\partial_\sigma A_\tau) +i \ell \cdot \oint d\sigma\, A_\sigma(\tau,\sigma)
\]
in the action,
where $y,\ell$ are Lagrange multipliers, and the flatness condition $\partial_\tau A_\sigma=\partial_\sigma A_\tau$ makes the holonomy $\oint d\sigma A_\sigma(\tau,\sigma)$ independent of $\tau$. 

To see that the new action is equivalent to the old one, we first fix the gauge as $x=0$ because the connection $A$ is flat, and then integrate the Lagrange multipliers that forces $A_\alpha=-\partial_\alpha x'$ for a function $x'$. Therefore, replacing $\partial_\alpha x$ with $\partial_\alpha x-A_\alpha |_{x=0}= \partial_\alpha x'$ in the bulk action is just a renaming of $x$.

On the other hand, we would like to integrate $A$ in the new action to get the T-dual action. We first fix the gauge $x=0$ and keep Lagrange multipliers, and the new action reads
\[
\begin{aligned}
S=&\frac{\sqrt{\lambda}}{4 \pi} \int_{\mathcal{D}} d \sigma d \tau \frac{(\partial_\alpha z)^2+A_\alpha^2}{z^2}-i \sum_{i=1}^n K_i \cdot \int^{\sigma_{i+1}}_{\sigma_i} d\sigma\,A_\sigma(0,\sigma)+i  q \cdot \int_{-\infty}^0 \!\!\!d\tau \,A_\tau(\tau,\gamma(\tau))\\
&+i\int_{\mathcal{D}} d \sigma d \tau\, y\cdot (\partial_\tau A_\sigma-\partial_\sigma A_\tau) +i \ell \cdot \oint d\sigma\, A_\sigma(-\infty,\sigma),
\end{aligned}
\]
where we fix $\tau=-\infty$ in the holonomy term since it is independent of $\tau$. Then integrating the term $y\cdot (\partial_\tau A_\sigma-\partial_\sigma A_\tau)$ by parts and rescaling fields $(A,y,z)\to\left(\frac{\sqrt{\lambda}}{4 \pi} A, \frac{4 \pi}{\sqrt{\lambda}} y, \frac{\sqrt{\lambda}}{4 \pi} z\right)$, we get that
\[
\begin{aligned}
\biggl(\frac{\sqrt{\lambda}}{4 \pi}\biggr)^{-1} S=&\int_{\mathcal{D}} d \sigma d \tau \biggl(\frac{(\partial_\alpha z)^2+A_\alpha^2}{z^2}-i(A_\sigma \cdot \partial_\tau y-A_\tau \cdot \partial_\sigma y)\biggr)\\
&-i\int_{-\infty}^0 \!\!\!d\tau \,A_\tau(\tau,\gamma(\tau))\cdot [y(\tau,\sigma_{n+1})-y(\tau,\sigma_{1})-q]\\
&-i \sum_{i=1}^n \int^{\sigma_{i+1}}_{\sigma_i} d\sigma\,A_\sigma(0,\sigma)\cdot (K_i-y(0,\sigma))\\
& -i  \oint d\sigma\, A_\sigma(-\infty,\sigma)\cdot (-\ell+y(-\infty,\sigma)).
\end{aligned}
\]
The gaussian integral of $A_\alpha$ in the bulk produces T-dual spacetime metric
\begin{equation}
\frac{(\partial_\alpha z)^2}{z^2}+z^2(\partial_\alpha y)^2 = \frac{(\partial_\alpha r)^2+(\partial_\alpha y)^2}{r^2}
\end{equation}
with a new radial AdS direction $r:=z^{-1}$, and integration of the other $A$ on the boundary implies conditions
\begin{equation}
\begin{aligned}
&y(\tau,\sigma_{1}+2\pi-0^+)=y(\tau,\sigma_{1})+q,&&\\
&y(0,\sigma)=K_i=\sum_{j=1}^ik_j
&&\text{ for $\sigma\in (\sigma_i,\sigma_{i+1})$},\\
&y(-\infty,\sigma)=\ell  &&\text{ for $\sigma\in (\sigma_{1},\sigma_{1}+2\pi-0^+)$}.
\end{aligned}
\end{equation}
The first two conditions give a periodic light-like polygon Wilson loop closed to the new AdS boundary $r\sim 0$, whose cusps are given by $\{K_i+kq\,:\, 1\leq i\leq n,k\in\mathbb Z\}$, and near the new AdS horizon $r>\!\!\!>0$, we also have infinite points $\ell+kq$ for $k\in \mathbb Z$. 

Before the T-duality, the string has zero winding number, so it should have zero momentum number after. Therefore, we need to project out all states with non-zero momentum number, which is realized by integrating over $\ell$, see~\cite{Ben-Israel:2018ckc}. The object is dual to infinite insertions near the AdS horizon $r>\!\!\!>0$ that depends on the operator associated with the form factor. 

For the form factor of $\operatorname{Tr}(\phi^2)$, it is natural to conjecture these points correspond to periodic insertions of the chiral self-dual on-shell Lagrangian \cite{CaronHuot:2010ek,Ben-Israel:2018ckc}
\begin{equation}
\mathcal{L}=\frac{1}{4} F_{\alpha \beta} F^{\alpha \beta}-\frac{1}{4} \phi_{A B}\left[\psi_\alpha^A, \psi^{\alpha B}\right]-\frac{1}{64}\left[\phi_{A B}, \phi_{C D}\right]\left[\phi^{A B}, \phi^{C D}\right],
\end{equation}
which is in the same $\frac12$-BPS supermultiplet and also needs to be integrated in the perturbation expansion around the self-dual Yang-Mills.
A small disadvantage here is that this expansion at weak coupling starts at one loop. For the higher-loop correction of the Wilson loop, they are given by the Lagrangian insertion formula \cite{CaronHuot:2010ek,Ben-Israel:2018ckc},
\[
W[\mathcal C_n]^{(L)}=\frac{1}{L!}\int_{z_i\sim z_i+q} dz_1\cdots dz_L\biggl\langle\prod_{n_1,\dots,n_L\in\mathbb Z}\frac{\mathcal L^{[n_1]}(z_1+n_1q)}{\lambda^2}\cdots \frac{\mathcal L^{[n_L]}(z_L+n_Lq)}{\lambda^2}W[\mathcal C_n]\biggr\rangle_0,
\]
where $\mathcal L^{[k]}(z+kq):=\mathcal L(z)$. 
These Lagrangian insertions are contracted at tree level $\lambda^0$ (the gluon field $A$ scales as $\lambda^{1/2}$), and only taking the contribution of one period to avoid overcounting.

\subsection{Gauge invariance and non-planar square roots}

We show in this subsection that the gauge invariance is protected by the infinite periodic sum, and non-planar square roots come from the path ordering of points on different edges in the same periodic image.

Since we know from the last subsection that the $q$-direction of the dual space is compact, the related ``momentum" should be discrete! It means that the momentum $k$ obeys $\exp(ik\cdot q)=0$ or $k\cdot q=2\pi n$ for $n\in \mathbb Z$, so the coordinate space propagator in the $R_\xi$-gauge now is
\begin{equation}
\begin{aligned}
\mathsf{D}_{\xi}^{\mu\nu}(x):\!&=\int \frac{d^{4-2\epsilon}k}{k^2+i 0^+}\left(\eta^{\mu \nu}-(1-\xi) \frac{k^\mu k^\nu}{k^2}\right)e^{-ik\cdot x}\sum_{n\in \mathbb Z}\delta(k\cdot q-2\pi n)\\
&=\int \frac{d^{4-2\epsilon}k}{k^2+i 0^+}\left(\eta^{\mu \nu}-(1-\xi) \frac{k^\mu k^\nu}{k^2}\right)e^{-ik\cdot x}\sum_{n\in \mathbb Z}e^{-ink\cdot q}\\
&=\sum_{n\in \mathbb Z}\int \frac{d^{4-2\epsilon}k}{k^2+i 0^+}\left(\eta^{\mu \nu}-(1-\xi) \frac{k^\mu k^\nu}{k^2}\right)e^{-ik\cdot (x+nq)}\\
&= \sum_{n=-\infty}^\infty D^{\mu\nu}_{\xi}(x+nq),
\end{aligned}
\end{equation}
where we use the Poisson summation formula to get the second line, 
and this is in fact the most natural way to make the propagator invariant under the discrete transformation $x\to x+q$, and we similarly define that 
\[
\mathsf{I}_\xi(i,j):=\int d(z_i)_\mu d(z_j)_\nu \mathsf{D}_{\xi}^{\mu\nu}(z_i-z_j)=\sum_{n\in \mathbb Z}I_\xi(i,j+n).
\]
In this note, we take the convention that a quantity denoted by the {\sf sans-serif} font is the infinite summed version of the quantity denoted by the serif font.

We first argue that the naive one-loop correction of the Wilson loop
\[
\frac12\sum_{i,j=1}^n \mathsf{I}_{\xi}(i,j)=
\frac12\lim_{N\to \infty}\sum_{j=-N}^N \sum_{i=1}^n I_\xi(i,j)
\]
is gauge-invariant! To see it, we use the box representation of $S_\xi(i,j)$ defined in Introduction, and we find that its $S_\xi$-contribution  
\[
\begin{aligned}
\lim_{N\to \infty}\sum_{j=-N}^N \sum_{i=1}^n S_\xi(i,j)&\propto \lim_{N\to\infty}\begin{tikzpicture}[scale=1,baseline={([yshift=-0.5ex]current bounding box.center)}]
\draw (0.5,-0.5) -- (0.5,-3) -- (3,-3) -- (3,-0.5) -- cycle;
\draw (0.5,-2.5) -- (3,-2.5);
\draw (0.5,-2) -- (3,-2);
\draw (0.5,-1.5) -- (3,-1.5);
\draw (0.5,-1) -- (3,-1);
\draw (1,-0.5) -- (1,-3);
\draw (1.5,-0.5) -- (1.5,-3);
\draw (2,-0.5) -- (2,-3);
\draw (2.5,-0.5) -- (2.5,-3);
\node at (0.75,-1.75) {\tiny$11$};
\node at (1.25,-1.75) {\tiny$21$};
\node at (1.75,-1.75) {\tiny$31$};
\node at (2.25,-1.75) {\tiny $~\cdots$};
\node at (2.75,-1.75) {\tiny $n1$};
\foreach \x in {0.75,1.25,...,2.75}
\node at (\x,-2.15) {\tiny $\vdots$};
\foreach \x in {0.75,1.25,...,2.75}
\node at (\x,-1.15) {\tiny $\vdots$};
\node at (0.75,-0.75) {\tiny$1N$};
\node at (1.25,-0.75) {\tiny$2N$};
\node at (1.75,-0.75) {\tiny$3N$};
\node at (2.25,-0.75) {\tiny $~\cdots$};
\node at (2.75,-0.75) {\tiny $nN$};
\node at (0.75,-2.75) {\tiny$1\,{\raisebox{1pt}{\scalebox{0.75}{-}}\!}N$};
\node at (1.25,-2.75) {\tiny$2\,{\raisebox{1pt}{\scalebox{0.75}{-}}\!}N$};
\node at (1.75,-2.75) {\tiny$3\,{\raisebox{1pt}{\scalebox{0.75}{-}}\!}N$};
\node at (2.25,-2.75) {\tiny $~\cdots$};
\node at (2.75,-2.75) {\tiny $n\,{\raisebox{1pt}{\scalebox{0.75}{-}}\!}N$};
\foreach \y in {-2.58,-1.58,-0.58}
{\foreach \x in {0.58, 1.08, 1.58, 2.08, 2.58}
\node at (\x,\y) {\tiny${\raisebox{1pt}{\scalebox{0.75}{-}}}$};
\foreach \x in {0.93, 1.43, 1.93, 2.43, 2.93}
\node at (\x,\y) {\tiny${\raisebox{1pt}{\scalebox{0.50}{+}}}$};
\foreach \x in {0.58, 1.08, 1.58, 2.08, 2.58}
\node at (\x,\y-0.37) {\tiny${\raisebox{1pt}{\scalebox{0.50}{+}}}$};
\foreach \x in {0.93, 1.43, 1.93, 2.43, 2.93}
\node at (\x,\y-0.37) {\tiny${\raisebox{1pt}{\scalebox{0.75}{-}}}$};};
\end{tikzpicture}\\
&= \lim_{N\to \infty}(
(-x_{1, -N}^2)^\epsilon + (-x_{n+1,N+1}^2)^\epsilon -(-x_{1, N+1}^2)^\epsilon - (-x_{n+1,-N}^2)^\epsilon )\\
&=\lim_{N\to \infty}\log\biggl(
\frac{x_{1,-N}^2 x_{n+1,N+1}^2}{x_{1,N+1}^2 x_{n+1,-N}^2}
\biggr)+O(\epsilon)
\end{aligned}
\]
goes to $0$ since 
\[
\lim_{N\to \infty} \frac{x_{1,-N}^2}{x_{n+1,-N}^2}= \lim_{x_{1,-N}\to\infty}\frac{x_{1,-N}^2}{x_{1,-N}^2-2q\cdot x_{1,-N}+q^2}=1
\]
and similarly $\lim_{N\to\infty} \log(x_{n+1,N+1}^2/x_{1,N+1}^2)=0$.
Therefore, the gauge dependent contribution vanishes when all edges are added, which saves the gauge invariance. However, this is not the correct one-loop correction of the Wilson loop, we should only consider diagrams that can be contained in one period~\cite{Brandhuber:2010ad}, which means we need to introduce a mechanism to project the infinite sum to a finite one. 
We here only use it as an example to show that we need to consider the infinite sum to keep the gauge invariance.

To see the correct way to get the one-loop expansion of the Wilson loop, we use the Lagrangian insertion formula. The one-loop correction is given by the diagram with one insertion with $\lambda^{-2}$, we need $4$ gluon fields $A$ which scales as $\lambda^{1/2}$, so the basic building block is 
\[
\int_{x_i}^{x_{i+1}} dx^{\gamma\dot\gamma}\int_{x_j}^{x_{j+1}} dy^{\sigma\dot\sigma}\int_{\ell\sim \ell+q} d^4\ell\sum_{n\in\mathbb Z}\langle \lambda^{-2}F_{\alpha\beta}(\ell+nq)F^{\alpha\beta}(\ell+nq)A_{\gamma\dot\gamma}(x)A_{\sigma\dot\sigma}(y)\rangle_0
\]
for points $x$ and $y$ on $\mathcal C_n$ are contained in one period, where
\begin{equation}
F_{\alpha\beta}=F_{\alpha\beta}=\frac12\epsilon^{\dot\alpha\dot\beta}F_{\alpha\dot\alpha,\beta\dot\beta}
\end{equation}
for $F_{\mu\nu}=\partial_\mu A_\nu-\partial_\nu A_\mu+\left[A_\mu, A_\nu\right]$. The infinite sum protects the gauge invariance, and they can be glued to an integral over the whole spacetime 
\[
\sum_{n\in \mathbb Z}\int_{\ell\sim\ell +q}d^4\ell \, O(\ell+nq;\dots)= \int_{\ell}d^4\ell \, O(\ell;\dots)
\]
for any function $O$. Then the integral 
\[
\int_{x_i}^{x_{i+1}} dx^{\gamma\dot\gamma}\int_{x_j}^{x_{j+1}} dy^{\sigma\dot\sigma}\int_{\ell} d^4\ell \langle \lambda^{-2}F_{\alpha\beta}(\ell)F^{\alpha\beta}(\ell)A_{\gamma\dot\gamma}(x)A_{\sigma\dot\sigma}(y)\rangle_0
\]
gives the one-loop chiral pentagon \cite{ArkaniHamed:2010gh,Caron-Huot:2011zgw} in the Feynman gauge, and it is non-trivial that it equals to $I_{\text{O}}(i,j)$. Then taking all contribution of pairs of edges in one period, we get the one-loop Wilson loop.

In the rest of this note, we do not use the Lagrangian insertion formula to compute the two-loop Wilson loop, but we still use the old-fashioned expansion of the bosonic Wilson loop in the Feynman gauge
\begin{equation}
\begin{aligned}
W[\mathcal{C}_n]= N_c^{-1}\operatorname{Tr}\biggl(&1+i g \int_{x\in \mathcal C_n} dx^\mu A_\mu^a(x) T^a -\frac{1}{2} g^2 \int_{x,y\in \mathcal C_n} dx^\mu(\lambda) dy^\nu(\tau) \\
& \times A_\mu^a(x) A_\nu^b(y)\left[T^a T^b \theta(\lambda-\tau)+T^b T^a \theta(\tau-\lambda)\right]+\cdots \biggr),
\end{aligned}
\end{equation}
because the diagrams are simpler, and most of them are computed in the literature, \textit{e.g.}~\cite{Anastasiou:2009kna,Drummond:2008aq}. However, we should remember that we need to first take the infinite periodic sum and then project diagrams that cannot be contained in one period.

An interesting consequence of this procedure is, if we consider two points on the same edges with the path ordering $z_1<z_2$, then in the infinite sum, there are diagrams where there is a path ordering between two points on different edges! We can draw it as
\begin{equation}
\begin{tikzpicture}[baseline={([yshift=-0.5ex]current bounding box.center)},scale=0.75]

\draw[thick] (-3.95,2.51) -- (-3.95,0.51);
\draw[thick] (-3.85,2.51) -- (-3.85,0.51);
\draw (-3.8,1.4) -- (-3.9,1.7) -- (-4,1.4);

\draw[thick] (1.3,2.5) -- (1.3,0.5);

\draw (1.4,1.4) -- (1.3,1.7) -- (1.2,1.4);

\node at (-4.4,2) {$z_2$};
\node at (-4.4,0.9) {$z_1$};
\node at (-3,1.4) {$\longrightarrow$};

\draw[decorate, decoration={brace, amplitude=5pt}] (0,0.3)--(-1,0.3);

\draw[thick] (-1,2.5) -- (-1,0.5);
\draw (-0.9,1.3) -- (-1,1.6) -- (-1.1,1.3);
\draw[thick] (0,2.5) -- (0,0.5);
\draw (0.1,1.3) -- (0,1.6) -- (-0.1,1.3);
\node at (-0.5,-0.2) {$q$};
      
 \node at (0.6,1.4) {$+$};
 
\draw[decorate, decoration={brace, amplitude=5pt}] (3.8,0.3)--(2.8,0.3);
\draw[thick] (2.8,2.5) -- (2.8,0.5);
\draw (2.9,1.3) -- (2.8,1.6) -- (2.7,1.3);
\draw[thick] (3.8,2.5) -- (3.8,0.5);
\draw (3.9,1.3) -- (3.8,1.6) -- (3.7,1.3);
\node at (3.3,-0.2) {$q$};

\node at (2,1.4) {$+$};
\node at (4.7,1.4) {$+\cdots$};
\node at (-1.9,1.4) {$\cdots+$};

\node[fill=blue,circle,inner sep=1.5pt] at (2.8,2) {};
\node[fill=red,circle,inner sep=1.5pt] at (3.8,0.9) {};
\node[fill=blue,circle,inner sep=1.5pt] at (1.3,2) {};
\node[fill=red,circle,inner sep=1.5pt] at (1.3,0.9) {};
\node[fill=blue,circle,inner sep=1.5pt] at (0,2) {};
\node[fill=red,circle,inner sep=1.5pt] at (-1,0.9) {};
\node[fill=blue,circle,inner sep=1.5pt] at (-3.9,2) {};
\node[fill=red,circle,inner sep=1.5pt] at (-3.9,0.9) {};
\end{tikzpicture}
\end{equation}
where the double line denotes that infinite sum of this edges, and the blue point is in front of the red point. 
This does not happen at one-loop level since all such diagrams $I_{\text{O}}(i,i+kn)\propto p_i^2=0$ for $k\in\mathbb Z$ vanish, and it starts to appear at two-loop level. 

At two-loop level, this explains how non-planar square roots appear.
Consider the following two-loop diagram,
\[
\begin{tikzpicture}[baseline={([yshift=-0.5ex]current bounding box.center)},scale=0.75]
\draw[thick] (-2.9,2.5) -- (-2.9,0.5);
\draw[thick] (-1.5,3.5) -- (1,3.5);
\draw[thick] (-2.8,2.5) -- (-2.8,0.5);
\draw[decorate,decoration={aspect=0.3,coil,segment length=4,amplitude=3}] 
        (-2.8,1) .. controls (-1.2,0.7) and (-0.5,1.5) .. (-0.5,3.5);
\node[fill=white,circle,inner sep=8pt] at (-0.87,1.4) {};
\draw[decorate,decoration={aspect=0.3,coil,segment length=4,amplitude=3}] 
        (-2.9,1.8) .. controls (-1.5,1) and (0.5,1.5) .. (0.5,3.5);
\node[fill=blue,circle,inner sep=2pt] at (-2.9,1.8) {};
\node[fill=red,circle,inner sep=2pt] at (-2.9,1) {};
\node at (-3.3,1) {$z_4$};
\node at (-3.3,1.8) {$z_1$};
\node at (-0.5,3.8) {$z_2$};
\node at (0.5,3.8) {$z_3$};
\end{tikzpicture}
\qquad \longrightarrow\qquad
\begin{tikzpicture}[baseline={([yshift=2ex]current bounding box.center)},scale=0.75]
\draw[thick] (-2.9,2.5) -- (-2.9,0.5);
\draw[thick] (-1.5,3.5) -- (1,3.5);
\draw[thick] (2.4,2.5) -- (2.4,0.5);
\draw[decorate,decoration={aspect=0.3,coil,segment length=4,amplitude=3}] 
        (-2.9,1.8) .. controls (-1.5,1) and (0.5,1.5) .. (0.5,3.5);
\node[fill=white,circle,inner sep=8pt] at (-0.1,2) {};
\draw[decorate,decoration={aspect=0.3,coil,segment length=4,amplitude=3}] 
        (2.4,1.1) .. controls (1,1) and (-0.5,1.5) .. (-0.5,3.5);
\node[fill=blue,circle,inner sep=2pt] at (-2.9,1.8) {};
\node[fill=red,circle,inner sep=2pt] at (2.4,1.1) {};
\draw[decorate, decoration={brace, amplitude=5pt}] (2.4,0.2)--(-2.9,0.2);    
\node at (-0.2,-0.3) {$q$};
\node at (-3.4,1.7) {$z_1$};
\node at (-0.5,3.8) {$z_2$};
\node at (0.5,3.8) {$z_3$};
\node at (3.2,1.1) {$z_4+q$};
\draw (-3,0.9) -- (-2.9,1.2) -- (-2.8,0.9);
\draw (2.3,1.8) -- (2.4,2.1) -- (2.5,1.8);
\end{tikzpicture}
\]
where the blue point are in front of the red point, we will explain in the next section why the diagram on the left hand side appears. The diagram on the right hand side is in its periodic sum with the path ordering between two points on different edges, and its integrand reads
\begin{equation}
\int_{z_4<z_1}\int_{z_2<z_3} (dz_1\cdot dz_3)(dz_2\cdot dz_4)  \,D(z_1-z_3)D(z_2-z_4-q).
\end{equation}
The full result of this integral is given in Section \ref{Xdiag}. It has one square root, which is exactly the one appearing in the non-planar Feynman diagrams with 4 massless and 1 massive legs \cite{Abreu:2023rco} like
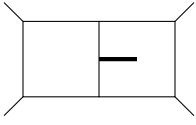
\begin{center}
\begin{tikzpicture}[scale=0.5]
\draw (-3.5,2) -- (-3.5,0) -- (-1.5,0) -- (-1.5,2) -- cycle;
\draw (-1.5,0) -- (0.5,0) -- (0.5,2) -- (-1.5,2);
\draw[ultra thick] (-1.5,1) -- (-0.5,1);
\draw (-4,2.5) -- (-3.5,2);
\draw (1,-0.5) -- (0.5,0);
\draw (-4,-0.5) -- (-3.5,0);
\draw (1,2.5) -- (0.5,2);
\end{tikzpicture}
\end{center}
Therefore, the non-planarity comes from the path ordering of points on different edges in the same periodic image!

\section{Two-loop Wilson loop diagrams}

In this section, we review the definition of the remainder functions of MHV form factors and discuss which Wilson loop diagrams we need to compute at the two-loop level by duality.

The remainder function $R_n$ of the $n$-particle form factor is defined by 
\begin{equation}
\mathcal F_n=\mathcal F_n^{\text{BDS}}\exp(R_n),
\end{equation}
where $\mathcal F_n := F_n/F^{\text{tree}}_n$ and the Bern-Dixon-Smirnov (BDS) ansatz~\cite{Bern:2005iz} $\mathcal F_n^{\text{BDS}}$ is almost the exponential of the one-loop form factor
\begin{equation}
\mathcal{F}_n^{\mathrm{BDS}}=\exp \left[\sum_{\ell=1}^{\infty} g^{2 \ell}\left(f^{(\ell)}(\epsilon) \mathcal{F}_n^{(1)}(\ell \epsilon)+C^{(\ell)}\right)\right],
\end{equation}
which captures all infrared divergence so that the remainder function $R_n$ is free of infrared divergence. 
At two-loop, the expansion reads
\begin{equation}\label{2Lremainder}
R_n^{(2)}=\mathcal{F}_n^{(2)}(\epsilon)-\frac{1}{2}\left(\mathcal{F}_n^{(1)}(\epsilon)\right)^2-f^{(2)}(\epsilon) \mathcal{F}_n^{(1)}(2 \epsilon)-C^{(2)}+\mathcal{O}(\epsilon),
\end{equation}
where 
\begin{equation}
f^{(2)}(\epsilon)=-2 \zeta_2-2 \zeta_3 \epsilon-2 \zeta_4 \epsilon^2+\mathcal{O}(\epsilon^3),\quad C^{(2)}=4\zeta_4.
\end{equation}
The same combination also appears at the expansion of the logarithmic of the form factor, that is 
\[
\log (\mathcal F_n) = \log(1+g^2\mathcal F_n^{(1)}+g^4\mathcal F_n^{(2)} + O(g^6)) = g^2\mathcal F_n^{(1)}+g^4(F_n^{(2)}-(\mathcal{F}_n^{(1)})^2/2) + O(g^6),
\]
so by the duality, we have that
\begin{equation}
\begin{aligned}
R_n^{(2)}&=\log (\mathcal F_n)^{(2)}-f^{(2)}(\epsilon) \mathcal{F}_n^{(1)}(2 \epsilon)-C^{(2)}+\mathcal{O}(\epsilon)\\
&=\log (\mathcal W[\mathcal{C}_n])^{(2)}-f^{(2)}(\epsilon) \mathcal{F}_n^{(1)}(2 \epsilon)-C^{(2)}+\mathcal{O}(\epsilon).
\end{aligned}
\end{equation}
Therefore, we are going to compute $\log (\mathcal W[\mathcal{C}_n])^{(2)}$ and see that its divergence is as mild as a one-loop form factor. If one is interested in the symbol of form factors, we have
\[
\mathcal S[R_n^{(2)}]=\mathcal S[\log(W[\mathcal{C}_n])^{(2)}]
\]
because all zeta values vanish at symbol level, and it should be free of divergence.

To compute the logarithmic of the Wilson loop, the \textit{non-abelian exponentiation theorem} \cite{Gatheral:1983cz,Frenkel:1984pz} is quite useful. It says that we only need to sum up the diagrams with the `maximal non-abelian color factor' $C_FC_A^{L-1}$ at $L$ loops in all diagrams, where $C_F=(N_c^2-1)/(2N_c)$ ($C_A=N_c$) is the Casimir operator for the fundamental (adjoint) representation. We list some two-loop gluon diagrams below, the contributing diagrams should have color factor $C_FC_A$:
\begin{equation}\label{colorfactor}
\begin{aligned}
\begin{tikzpicture}[baseline={([yshift=-3.5ex]current bounding box.center)},scale=0.5]
\draw[decorate,decoration={aspect=0.3,coil,segment length=4,amplitude=3}] 
        (-2.3736,-0.038) .. controls (-1.1523,0.1857) and (-0.2742,0.4456) .. (-0.7544,1.7753);
\draw[decorate,decoration={aspect=0.3,coil,segment length=4,amplitude=3}] 
        (2.8274,0.0623) .. controls (2.0964,0.3353) and (1.0899,-0.1758) .. (0.9472,1.8844);
\node at (-2.7765,-0.0649) {$a$};
\node at (-0.9585,2.0337) {$a$};
\node at (1.2085,2.1325) {$b$};
\node at (3.1212,0.1043) {$b$};
\draw (-2.5,-1) .. controls (-2.5,3) and (3,3) .. (3,-1);
\end{tikzpicture}
&\qquad N_c^{-1}\operatorname{Tr}(T^a T^a T^b T^b)=C_F^2\to 0, \\
\begin{tikzpicture}[baseline={([yshift=-3.5ex]current bounding box.center)},scale=0.5]
\draw[decorate,decoration={aspect=0.3,coil,segment length=4,amplitude=3}] 
        (-2.3736,-0.038) .. controls (-1.1523,0.1857) and (0.846,0.4728) .. (1.7393,1.5671);
\node[fill=white,circle,inner sep=8.5pt] at (0.3,0.6) {};
\draw[decorate,decoration={aspect=0.3,coil,segment length=4,amplitude=3}] 
        (2.8274,0.0623) .. controls (1.0007,0.0298) and (-0.7105,1.0791) .. (-1.334,1.5281);
\node at (-2.7765,-0.0649) {$a$};
\node at (-1.3846,1.901) {$b$};
\node at (1.9821,1.8412) {$a$};
\node at (3.1212,0.1043) {$b$};
\draw (-2.5,-1) .. controls (-2.5,3) and (3,3) .. (3,-1);
\end{tikzpicture}
&\qquad  
N_c^{-1}\operatorname{Tr}(T^a T^b T^a T^b)=C_F^2-\frac12 C_F C_A \to -\frac12 C_F C_A,\\
\begin{tikzpicture}[baseline={([yshift=-3.5ex]current bounding box.center)},scale=0.5]
\draw[decorate,decoration={aspect=0.3,coil,segment length=4,amplitude=3}] 
        (-2.3736,-0.038) .. controls (-1.5,-0.5) and (-0.5,-1) .. (0.5,-0.5);
\draw[decorate,decoration={aspect=0.3,coil,segment length=4,amplitude=3}] 
        (2.8274,0.0623) .. controls (2,-1) and (1,-0.5) .. (0.5,-0.5);
\draw[decorate,decoration={aspect=0.3,coil,segment length=4,amplitude=3}] 
        (0,2) .. controls (0,1) and (0,0) .. (0.5,-0.5);
\node at (-2.7765,-0.0649) {$a$};
\node at (0,2.5) {$b$};
\node at (3.1212,0.1043) {$c$};
\draw (-2.5,-1) .. controls (-2.5,3) and (3,3) .. (3,-1);
\end{tikzpicture}
&\qquad 
-if^{abc}N_c^{-1}\operatorname{Tr}(T^aT^bT^c)=\frac12 C_FC_A \to \frac12 C_FC_A,
\end{aligned}
\end{equation}
and the theorem says that the first diagram does not contribute, and the color factor of the second diagram is replaced to $-C_FC_A/2$. In fact, this is a product of two 1-loop diagrams, and it is already deleted in the expansion $\langle W[\mathcal C_n]\rangle^{(2)}-\frac12(\langle W[\mathcal C_n]\rangle^{(1)})^2$. In the large-$N_c$ limit, the two-loop color factor $C_F C_A$ combining with $g_{\text{YM}}^4$ becomes the square of ’t Hooft coupling
\[
C_FC_Ag_{\text{YM}}^4=\frac{N_c^2-1}{2}g_{\text{YM}}^4\sim g^4.
\]

At two-loop level, the self-energy diagram also has the color factor $C_FC_A$, but it does not contribute to the finite part of the remainder function~\cite{Anastasiou:2009kna}. We will discuss it together with Y diagrams in Section \ref{Ydiag}. In this note, we only consider last two topologies which contributes to the symbol, so there are six kinds of diagrams we need to compute:
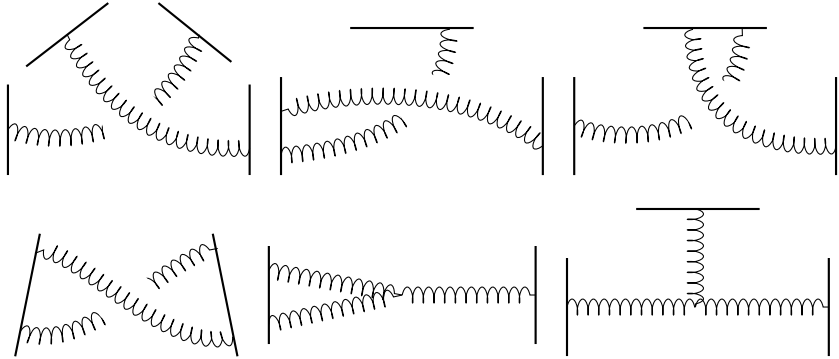
\[
\begin{tikzpicture}[scale=0.60,baseline={([yshift=-1.5ex]current bounding box.center)}]
\draw[thick] (-2.9,2.5) -- (-2.9,0.5);
\draw[thick] (-2.5,2.9) -- (-0.7,4.3);
\draw[thick] (2.4,2.5) -- (2.4,0.5);
\draw[decorate,decoration={aspect=0.3,coil,segment length=4,amplitude=3}] 
        (-2.9,1.5) .. controls (-1.5,1) and (0.5,1.5) .. (1.3,3.6);
\node[fill=white,circle,inner sep=8pt] at (-0.2,1.8) {};
\draw[decorate,decoration={aspect=0.3,coil,segment length=4,amplitude=3}] 
        (2.4,1.1) .. controls (1,1) and (-0.5,1.5) .. (-1.6,3.6);
\draw[thick] (2,3) -- (0.4,4.3);
\end{tikzpicture}
\quad 
\begin{tikzpicture}[scale=0.65,baseline={([yshift=-0.5ex]current bounding box.center)}]
\draw[thick] (-2.9,2.5) -- (-2.9,0.5);
\draw[thick] (-1.5,3.5) -- (1,3.5);
\draw[thick] (2.4,2.5) -- (2.4,0.5);
\draw[decorate,decoration={aspect=0.3,coil,segment length=4,amplitude=3}] 
        (-2.9,0.9) .. controls (-1.5,1) and (0.5,1.5) .. (0.5,3.5);
\node[fill=white,circle,inner sep=7.5pt] at (0,2) {};
\draw[decorate,decoration={aspect=0.3,coil,segment length=4,amplitude=3}] 
        (2.4,1.1) .. controls (1.2,2.1) and (-1,2.4) .. (-2.9,1.8);
\end{tikzpicture} 
\quad 
\begin{tikzpicture}[scale=0.65,baseline={([yshift=-0.5ex]current bounding box.center)}]
\draw[thick] (-2.9,2.5) -- (-2.9,0.5);
\draw[thick] (-1.5,3.5) -- (1,3.5);
\draw[thick] (2.4,2.5) -- (2.4,0.5);
\draw[decorate,decoration={aspect=0.3,coil,segment length=4,amplitude=3}] 
        (-2.9,1.5) .. controls (-1.5,1) and (0.5,1.5) .. (0.5,3.5);
\node[fill=white,circle,inner sep=7.5pt] at (-0.1,1.9) {};
\draw[decorate,decoration={aspect=0.3,coil,segment length=4,amplitude=3}] 
        (2.4,1.1) .. controls (1,1) and (-0.5,1.5) .. (-0.5,3.5);
\end{tikzpicture}
\]
\[
\begin{tikzpicture}[scale=0.65,baseline={([yshift=-0.5ex]current bounding box.center)}]
\draw[thick] (-2.5,3.5) -- (-3,1);
\draw[thick] (1.5,1) -- (1,3.5);
\draw[decorate,decoration={aspect=0.3,coil,segment length=4,amplitude=3}] 
        (-2.9,1.5) .. controls (-1.5,1) and (0,3) .. (1.1,3.2);
\node[fill=white,circle,inner sep=8pt] at (-0.8,2.2) {};
\draw[decorate,decoration={aspect=0.3,coil,segment length=4,amplitude=3}] 
        (1.4,1.4) .. controls (0,1) and (-2,3.5) .. (-2.6,3.1);
\end{tikzpicture}
\quad 
\begin{tikzpicture}[scale=0.65,baseline={([yshift=-0.5ex]current bounding box.center)}]
\draw[thick] (-3,2.5) -- (-3,0.5);
\draw[thick] (2.4,2.5) -- (2.4,0.5);
\draw[decorate,decoration={aspect=0.3,coil,segment length=4,amplitude=3}] (-3,0.9) -- (-0.3,1.5);
\draw[decorate,decoration={aspect=0.3,coil,segment length=4,amplitude=3}] (-0.3,1.5) -- (2.4,1.5);
\draw[decorate,decoration={aspect=0.3,coil,segment length=4,amplitude=3}] (-3,2) -- (-0.3,1.5);
\end{tikzpicture}
\quad 
\begin{tikzpicture}[scale=0.65,baseline={([yshift=-1.5ex]current bounding box.center)}]
\draw[thick] (-2.9,2.5) -- (-2.9,0.5);
\draw[thick] (-1.5,3.5) -- (1,3.5);
\draw[thick] (2.4,2.5) -- (2.4,0.5);
\draw[decorate,decoration={aspect=0.3,coil,segment length=4,amplitude=3}] (-2.9,1.5) -- (-0.3,1.5);
\draw[decorate,decoration={aspect=0.3,coil,segment length=4,amplitude=3}] (-0.3,1.5) -- (2.4,1.5);
\draw[decorate,decoration={aspect=0.3,coil,segment length=4,amplitude=3}] (-0.3,3.5) -- (-0.3,1.5);
\end{tikzpicture}
\]
other diagrams are their flip and periodic images. 
We should remember that we need to first take their infinite periodic sum and then project diagrams that cannot be contained in one period.

\section{Analysis of diagrams}\label{sec:diags}

In this section, we discuss the integrands and some results of the diagrams shown in the last section. We first consider how to compute a diagram with IR divergence from a generic diagram, and then we focus on generic diagrams.

\paragraph{Regularization}

Instead of dimensional regularization, we would like to use a regularization which is more compatible with the integration over the Wilson line. Since almost all IR divergence (or UV divergence in the sense of Wilson loop) appears (in the Feynman gauge) when the diagram contains cusps, one can first compute the generic diagram without cusp, which is finite and independent of any regularization, and then take the following limit to diagrams with cusps. At each cusp, we introduce a small cutoff as 
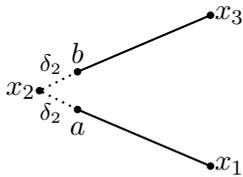
\[
\begin{tikzpicture}[scale=0.5]
\draw[thick] (0,4) -- (3.5,5.5);
\draw[thick] (0,3) -- (3.5,1.5);
\draw[thick,dotted] (0,4) -- (-1,3.5);
\draw[thick,dotted] (0,3) -- (-1,3.5);
\node at (-0.7,2.9) {\footnotesize $\delta_{2}$};
\node at (-0.7,4.2) {\footnotesize $\delta_{2}$};
\node at (4,1.5) {$x_1$};
\node at (-1.5,3.5) {$x_2$};
\node at (4,5.5) {$x_3$};
\node at (0,4.5) {$b$};
\node at (0,2.5) {$a$};
\node[circle,inner sep=1pt,fill=black] at (-1,3.5) {};
\node[circle,inner sep=1pt,fill=black] at (3.5,1.5) {};
\node[circle,inner sep=1pt,fill=black] at (3.5,5.5) {};
\node[circle,inner sep=1pt,fill=black] at (0,3) {};
\node[circle,inner sep=1pt,fill=black] at (0,4) {};
\end{tikzpicture}
\]
where 
\begin{equation}\label{higgsreg}
a=(1-\delta_{2})x_2+\delta_{2} x_1,\quad 
b=(1-\delta_{2})x_2+\delta_{2} x_3,
\end{equation}
then the cusp $a$ becomes a massive interval $a-b$ with a mass $\delta_{2}^2 x_{13}^2$ and massless edges get the mass as 
\[
(x_1-b)^2=(x_3-a)^2=\delta_{2}(x_1-x_3)^2=\eta,
\]
and so we set $\delta_{2}=\eta/x_{13}^2$
for this cusp. This regularization is equivalent to saying that each massless edge has a small mass $\eta$, which makes the regularization uniform for all cusps. Then one can get the result of a divergent diagram depending on $x_1,x_2,x_3$ by taking the $\eta\to 0^+$ limit to the generic diagram depending on dual points $a,b,x_1,x_3$.

In \cite{Drummond:2007au,Drummond:2008aq},
two ancillary  divergent diagrams are introduced to capture the divergent part of the other diagrams in dimensional regularization. We here introduce the same diagrams, but in our regularization. We define a scalar three-point vertex
\begin{equation}
T_\epsilon(z_1,z_2,z_3) = \hspace{-2.5ex}\begin{tikzpicture}[scale=0.5,baseline={([yshift=-0.5ex]current bounding box.center)}]
\draw (-2,0) -- (-0.5,1.5);
\draw (-0.5,1.5) -- (1.5,1.5);
\draw (-2,3) -- (-0.5,1.5);
\node[fill=black,circle,inner sep=2pt] at (-0.5,1.5) {};
\node at (-2.5,0) {$z_1$};
\node at (-2.5,3) {$z_2$};
\node at (2,1.5) {$z_3$};
\end{tikzpicture} = \Gamma(1-2\epsilon)\int_{\mathbb R^3_+} d^3a \frac{ (a_1a_2a_3)^{-\epsilon}\,\delta(a_1+a_2+a_3-1)}{\left(\frac12\sum_{i,j=1}^3a_ia_j(z_i-z_j)^2\right)^{1-2\epsilon}},
\end{equation}
and ancillary  diagrams are defined in this vertex in both dimensional regularization and our cutoff regularization as
\begin{align}\label{Mp}
M^{+}_{ij}&:=
\begin{tikzpicture}[scale=0.5,baseline={([yshift=-0.5ex]current bounding box.center)}]
\draw[thick] (-3,2.5) -- (-3,0.5);
\draw[thick] (2.5,2.5) -- (2.5,0.5);
\draw (-3,0.5) -- (-0.3,1.5);
\draw (-0.3,1.5) -- (2.5,1.5);
\draw (-3,1.9) -- (-0.3,1.5);
\node[fill=black,circle,inner sep=2pt] at (-0.3,1.5) {};
\node at (-3.5,0) {$x_{i}$};
\node at (-3.5,3) {$x_{i+1}$};
\node at (3,3) {$x_j$};
\node at (3,0) {$x_{j+1}$};
\node at (-3.5,2) {$z_i$};
\node at (3,1.5) {$z_j$};
\end{tikzpicture}= \int_0^1 dt_j\int_{0}^1 dt_i \,(p_i\cdot p_j) \, T_\epsilon(x_i,z_i,z_j)\\
&\xrightarrow{\text{regularize}}
\int_0^1 dt_j\int_{\delta_{i}}^1 dt_i \,(p_i\cdot p_j)\, T_0(x_i-\delta_{i}p_{i-1},\,z_i,\,z_j),\nonumber\\
\label{Mm}
M^{-}_{ij}&:=
\begin{tikzpicture}[scale=0.5,baseline={([yshift=-0.5ex]current bounding box.center)}]
\draw[thick] (-3,2.5) -- (-3,0.5);
\draw[thick] (2.5,2.5) -- (2.5,0.5);
\draw (-3,1.1) -- (-0.3,1.5);
\draw (-0.3,1.5) -- (2.5,1.5);
\draw (-3,2.5) -- (-0.3,1.5);
\node[fill=black,circle,inner sep=2pt] at (-0.3,1.5) {};
\node at (-3.5,0) {$x_{i-1}$};
\node at (-3.5,3) {$x_{i}$};
\node at (3,3) {$x_j$};
\node at (3,0) {$x_{j+1}$};
\node at (-4,1) {$z_{i{-}1}$};
\node at (3,1.5) {$z_j$};
\end{tikzpicture}=\int_0^1 dt_j\int^{1}_0 dt_{i-1}\, (p_{i-1}\cdot p_j) \, T_\epsilon(z_{i-1},x_i,z_j)\\
&\xrightarrow{\text{regularize}}
\int_0^1 dt_j\int_{0}^{1-\delta_{i}} dt_{i-1} \,(p_{i-1}\cdot p_j)\, T_0(z_{i-1},\,x_i+\delta_{i}p_i,\,z_j),\nonumber
\end{align}
where $\delta_{i}:=\eta/x_{i-1,i+1}^2$. Our regularization is simply moving the point at the cusp to the adjacent edges with a small shift $\delta_{i}$ and introducing the same cutoff to the Wilson line integral.

The finite $T_0(x_1,x_2,x_3)$ vertex defined above is the one-loop three-mass triangle integral in 4D
\[
T_0(x_1,x_2,x_3) \propto \int  \frac{d^4 x}{(x-x_1)^2(x-x_2)^2(x-x_3)^2},
\]
whose result is 
\begin{equation}
T_0(x_1,x_2,x_3) =\frac{2}{(x_1-x_3)^2(z-\bar z)}\biggl(\mathrm{Li}_2(z)-\mathrm{Li}_2(\bar{z})+\frac{1}{2} \log (z \bar{z}) \log \left(\frac{1-z}{1-\bar{z}}\right)
\biggr),
\end{equation}
where $z$ and $\bar z$ are defined by equations
\[
z\bar z=\frac{(x_1-x_2)^2}{(x_1-x_3)^2},\quad 
(1-z)(1-\bar z)=\frac{(x_2-x_3)^2}{(x_1-x_3)^2},
\]
or manifestly
\begin{equation}\label{zzbar}
z,\bar z=\frac{x_{12}^2+x_{13}^2-x_{23}^2\pm\sqrt{(x_{12}^2+x_{13}^2-x_{23}^2)^2-4 x_{12}^2 x_{13}^2}}{2 x_{13}^2}.
\end{equation}

It is easy to see from the definition that 
\[
M_{ij}^+=-M_{ij}^-(x_{i+1}\leftrightarrow x_{i-1}),
\]
and it has the expansion for $j\neq i,i+1$
\begin{equation}
M_{ij}^+=M_{ij}^{+,(-1)}\epsilon^{-1}+O(1)=M_{ij}^{+,(-1)}\log(\eta)+O(1)
\end{equation}
in two regularization schemes. They share the same weight-3 polylogarithm $M_{ij}^{+,(-1)}$ as the coefficient of divergence, but their $O(1)$ parts are different.
We will see in the following subsections that these two diagrams exactly capture the non-trivial divergent parts of all the other diagrams.

\subsection{One-loop square diagrams}

The following diagram is just the product of two one-loop diagram:
\[
I_{\text{O}}(i,k)I_{\text{O}}(j,l)=\begin{tikzpicture}[scale=0.60,baseline={([yshift=-0.5ex]current bounding box.center)}]
\draw[thick] (-2.9,2.5) -- (-2.9,0.5);
\draw[thick] (-2.5,2.9) -- (-0.7,4.3);
\draw[thick] (2.4,2.5) -- (2.4,0.5);
\draw[decorate,decoration={aspect=0.3,coil,segment length=4,amplitude=3}] 
        (-2.9,1.5) .. controls (-1.5,1) and (0.5,1.5) .. (1.3,3.6);
\node[fill=white,circle,inner sep=7.2pt] at (-0.2,1.8) {};
\draw[decorate,decoration={aspect=0.3,coil,segment length=4,amplitude=3}] 
        (2.4,1.1) .. controls (1,1) and (-0.5,1.5) .. (-1.6,3.6);
\node at (-3.5,1.5) {$z_i$};
\node at (-1.9,4) {$z_j$};
\node at (1.5,3.9) {$z_k$};
\node at (3,1.1) {$z_l$};
\draw[thick] (2,3) -- (0.4,4.3);
\end{tikzpicture}
:=\int\frac{2(dz_i\cdot dz_k)}{(z_i-z_k)^2}\int\frac{2(dz_j\cdot dz_l)}{(z_j-z_l)^2},
\]
and it always converges. The result of $I_{\text{O}}$ is a weight-two polylogarithm, 
\begin{align}
I_{\text{O}}(i,k):=&\int\frac{2(dz_i\cdot dz_k)}{(z_i-z_k)^2}=\int_{[0,1]^2}\!\!\!\! dt_i dt_k\frac{(p_i+p_k)^2}{((x_i(1-t_i)+x_{i+1}t_i)-(x_k(1-t_k)+x_{k+1}t_k))^2}\nonumber\\
=&\,\operatorname{Li}_2\biggl(1-\frac{(x_{i,k}^2-x_{i,k+1}^2) (x_{i,k+1}^2-x_{i+1,k+1}^2)}{x_{i,k+1}^2 x_{i+1,k}^2-x_{i,k}^2 x_{i+1,k+1}^2}\biggr)\\
&+\log (x_{i,k+1}^2) \log \biggl(\frac{(x_{i,k}^2-x_{i,k+1}^2) (x_{i,k+1}^2-x_{i+1,k+1}^2)}{x_{i,k+1}^2 x_{i+1,k}^2-x_{i,k}^2 x_{i+1,k+1}^2}\biggr)\nonumber\\
&-(i\leftrightarrow i+1)-(k\leftrightarrow k+1)
+(i\leftrightarrow i+1, k\leftrightarrow k+1),\nonumber
\end{align}
and in fact it is also the finite part of a two-mass easy box function, see \textit{e.g.}~\cite{Brandhuber:2010ad}.

\subsection{Curtain diagrams}

For curtain diagrams, we only need to consider the following three diagrams:
\begin{equation}
I_{\text{C},1}(i,j,k)=
\begin{tikzpicture}[scale=0.65,baseline={([yshift=-1.5ex]current bounding box.center)}]
\draw[thick] (-2.9,2.5) -- (-2.9,0.5);
\draw[thick] (-1.5,3.5) -- (1,3.5);
\draw[thick] (2.4,2.5) -- (2.4,0.5);
\draw[decorate,decoration={aspect=0.3,coil,segment length=4,amplitude=3}] 
        (-2.9,0.9) .. controls (-1.5,1) and (0.5,1.5) .. (0.5,3.5);
\node[fill=white,circle,inner sep=7.2pt] at (0,2) {};
\draw[decorate,decoration={aspect=0.3,coil,segment length=4,amplitude=3}] 
        (2.4,1.1) .. controls (1.2,2.1) and (-1,2.4) .. (-2.9,1.8);
\node at (-3.3,0.9) {$z_1$};
\node at (-3.3,1.8) {$z_2$};
\node at (0.5,3.8) {$z_3$};
\node at (2.8,1) {$z_4$};
\node at (-3.2,0.2) {$x_i$};
\node at (-3,2.8) {$x_{i+1}$};
\node at (-1.6,3.8) {$x_j$};
\node at (1.4,3.8) {$x_{j+1}$};
\node at (2.4,2.8) {$x_k$};
\node at (2.6,0.2) {$x_{k+1}$};
\end{tikzpicture} =4\int_{z_1<z_2} \frac{(dz_1\cdot dz_3)(dz_2\cdot dz_4)}{(z_1-z_3)^2(z_2-z_4)^2},
\end{equation}
\begin{equation}
I_{\text{C},2}(i,j,k)=\begin{tikzpicture}[scale=0.65,baseline={([yshift=-1.5ex]current bounding box.center)}]
\draw[thick] (-2.9,2.5) -- (-2.9,0.5);
\draw[thick] (-1.5,3.5) -- (1,3.5);
\draw[thick] (2.4,2.5) -- (2.4,0.5);
\draw[decorate,decoration={aspect=0.3,coil,segment length=4,amplitude=3}] 
        (-2.9,1.5) .. controls (-1.5,1) and (0.5,1.5) .. (0.5,3.5);
\node[fill=white,circle,inner sep=7.3pt] at (0,2) {};
\draw[decorate,decoration={aspect=0.3,coil,segment length=4,amplitude=3}] 
        (2.4,1.1) .. controls (1,1) and (-0.5,1.5) .. (-0.5,3.5);
\node at (-3.2,1.5) {$z_1$};
\node at (-0.5,3.8) {$z_2$};
\node at (0.5,3.8) {$z_3$};
\node at (2.8,1.1) {$z_4$};
\node at (-3.2,0.2) {$x_i$};
\node at (-3,2.8) {$x_{i+1}$};
\node at (-1.6,3.8) {$x_j$};
\node at (1.4,3.8) {$x_{j+1}$};
\node at (2.4,2.8) {$x_k$};
\node at (2.6,0.2) {$x_{k+1}$};
\end{tikzpicture}
= 4\int_{z_2<z_3} \frac{(dz_1\cdot dz_3)(dz_2\cdot dz_4)}{(z_1-z_3)^2(z_2-z_4)^2},
\end{equation}
\begin{equation}
I_{\text{C},3}(i,j,k)=
\begin{tikzpicture}[scale=0.65,baseline={([yshift=-1.5ex]current bounding box.center)}]
\draw[thick] (-2.9,2.5) -- (-2.9,0.5);
\draw[thick] (-1.5,3.5) -- (1,3.5);
\draw[thick] (2.4,2.5) -- (2.4,0.5);
\draw[decorate,decoration={aspect=0.3,coil,segment length=4,amplitude=3}] 
        (2.4,2) .. controls (1.2,2.1) and (-1,2.4) .. (-2.9,1);
\node[fill=white,circle,inner sep=7pt] at (0,2) {};
\draw[decorate,decoration={aspect=0.3,coil,segment length=4,amplitude=3}] 
        (2.4,1) .. controls (1.5,1.5) and (0,1) .. (-0.5,3.5);
\node at (-3.3,0.9) {$z_1$};
\node at (-0.5,3.8) {$z_2$};
\node at (2.8,2) {$z_3$};
\node at (2.8,1) {$z_4$};
\node at (-3.2,0.2) {$x_i$};
\node at (-3,2.8) {$x_{i+1}$};
\node at (-1.6,3.8) {$x_j$};
\node at (1.4,3.8) {$x_{j+1}$};
\node at (2.4,2.8) {$x_k$};
\node at (2.6,0.2) {$x_{k+1}$};
\end{tikzpicture}
=
4\int_{z_3<z_4} \frac{(dz_1\cdot dz_3)(dz_2\cdot dz_4)}{(z_1-z_3)^2(z_2-z_4)^2},
\end{equation}
where $i<j<k$.
It is straightforward to integrate them to weight-4 polylogarithms, whose letters are rational functions in planar variables $x_{ab}^2$.

When the above diagrams contain cusps, they have IR divergences, and we find that they are captured by the ancillary  diagrams as
\begin{equation}\label{divC}
\begin{aligned}
&I_{\text{C,1}}(i-1,i,k)=2 M^-_{i,k}+O(1),&&\hspace{-6ex}I_{\text{C,2}}(i-1,i,k)=2 M^+_{i,k}+O(1),\\
&I_{\text{C,2}}(i-1,k-1,k)=2 M^-_{k,i-1}+O(1),&&\hspace{-6ex}I_{\text{C,3}}(i-1,k-1,k)=2 M^+_{k,i-1}+O(1),\\
&I_{\text{C,2}}(1,2,3)=
\biggl(\frac{\zeta _2}{\epsilon ^2}-\frac{2 \zeta _3}{\epsilon }+\frac{19 }{4}\zeta _4\biggr)(x_{13}^2x_{24}^2)^{\epsilon}
,&&\hspace{-10ex}
\end{aligned}
\end{equation}
where $k\neq i+1$. The divergence of $I_{\text{C,2}}(1,2,3)$ will be absorbed into the one-loop part in the eq.\eqref{2Lremainder}, and the other $O(1)$ parts are independent of regularization schemes.

\subsection{Star diagrams (Hard diagrams)}\label{sec:star}

For the star diagram that are called hard diagrams in \cite{Anastasiou:2009kna}, 
\[
I_{\text{S}}(i,j,k)=
\begin{tikzpicture}[scale=0.75,baseline={([yshift=-2.5ex]current bounding box.center)}]
\draw[thick] (-2.9,2.5) -- (-2.9,0.5);
\draw[thick] (-1.5,3.5) -- (1,3.5);
\draw[thick] (2.4,2.5) -- (2.4,0.5);
\draw[decorate,decoration={aspect=0.3,coil,segment length=4,amplitude=3}] (-2.9,1.5) -- (-0.3,1.5);
\draw[decorate,decoration={aspect=0.3,coil,segment length=4,amplitude=3}] (-0.3,1.5) -- (2.4,1.5);
\draw[decorate,decoration={aspect=0.3,coil,segment length=4,amplitude=3}] (-0.3,3.5) -- (-0.3,1.5);
\node at (-3.5,1.5) {$z_i$};
\node at (-0.25,4) {$z_j$};
\node at (3,1.5) {$z_k$};
\node at (-3.2,0.2) {$x_i$};
\node at (-3,2.8) {$x_{i+1}$};
\node at (-1.6,3.8) {$x_j$};
\node at (1.2,3.8) {$x_{j+1}$};
\node at (2.4,2.8) {$x_k$};
\node at (2.6,0.2) {$x_{k+1}$};
\end{tikzpicture}
\]
we define 
\[
z_a=(1-t_a)x_a+t_ax_{a+1}\quad \text{for $a=i,j,k$ and $0\leq t_a\leq 1$},
\]
and the momentum $p_a:=x_{a+1}-x_a$ is massless, the gluon three-point vertex gives a numerator
\[
V=(dz_i\cdot dz_j) (dz_k\cdot (\partial_i-\partial_j))
+(dz_j\cdot dz_k) (dz_i\cdot (\partial_j-\partial_k))
+(dz_k\cdot dz_i) (dz_j\cdot (\partial_k-\partial_i)),
\]
where $\partial_a:=\partial/\partial z_a$, and the integrand is 
\begin{equation}\label{IS}
\begin{aligned}
I_{\text{S}}(i,j,k)&= \int_{z_i,z_j,z_k} V\, T_\epsilon (z_i,z_j,z_k)\\
&= \Gamma(1-2\epsilon)\int_{z_i,z_j,z_k}V\int_{\mathbb R^3_+} d^3a \frac{(a_1a_2a_3)^{-\epsilon}\delta(a_1+a_2+a_3-1)}{\left(\frac12\sum_{i,j=1}^3a_ia_j(z_i-z_j)^2\right)^{1-2\epsilon}}.
\end{aligned}
\end{equation}

The $(3+3)$-fold integration of the star diagram eq.\eqref{IS} is tedious but straightforward. One can first integrate $a_1,a_2,t_2$ and $t_3$, then the integrand for $a_3$ and $t_1$ is still free of square roots, and the integration of $a_3$ introduces square roots
depending on $t_1$. A similar Wilson line integral with square roots is already considered for the Wilson loop $d\log$ representation of generic two-loop double pentagons \cite{He:2020lcu,He:2020uxy}. We follow the same method here to integrate the remaining $t_1$-integral by rationalization.

Since there are $2^3=8$ vertices for the integration domain $\{t_a\in [0,1]\}_{a=i,j,k}$, the result of $I_{\text{S}}$ has $8$ one-loop triangle square roots eq.\eqref{zzbar}. For generic kinematics, there are $7$ algebraic letters for each square root in this symbol. These letters and the full result are too lengthy, so we leave them in the ancillary  files. 

If the star diagram contains a cusp, it has the IR divergence\footnote{This is gauge dependent, and it is free of divergence in the axial gauge, see \cite{Drummond:2007au}.}. Consider the following diagram,
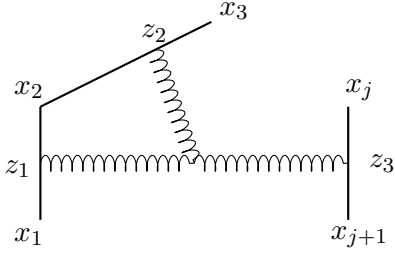
\[
\begin{tikzpicture}[scale=0.75]
\draw[thick] (-3,2.5) -- (-3,0.5);
\draw[thick] (-3,2.5) -- (0,4);
\draw[thick] (2.4,2.5) -- (2.4,0.5);
\draw[decorate,decoration={aspect=0.3,coil,segment length=4,amplitude=3}] (-3,1.5) -- (-0.3,1.5);
\draw[decorate,decoration={aspect=0.3,coil,segment length=4,amplitude=3}] (-0.3,1.5) -- (2.4,1.5);
\draw[decorate,decoration={aspect=0.3,coil,segment length=4,amplitude=3}] (-1,3.5) -- (-0.3,1.5);
\node at (-3.4,1.4) {$z_1$};
\node at (-1,3.8) {$z_2$};
\node at (3,1.5) {$z_3$};
\node at (-3.2,0.2) {$x_1$};
\node at (-3.2,2.8) {$x_2$};
\node at (0.4,4.2) {$x_3$};
\node at (2.6,2.8) {$x_j$};
\node at (2.6,0.2) {$x_{j+1}$};
\end{tikzpicture}
\]
since $(z_1-z_2)^2=(1-t_1)t_2(x_1-x_3)^2$ in this case, the divergent region is given by $a_3\sim 0$, where
\[
T_0(z_1,z_2,z_3)\propto \frac{1}{(1-t_1)t_2(x_1-x_3)^2},
\]
so it diverges when $t_1\sim 1$ or $t_2\sim 0$. To capture its divergence, we introduce another numerator
\begin{equation}
V_{\text{div}}:=(dz_2\cdot dz_3) (dz_1\cdot \partial_1)
-(dz_3\cdot dz_1) (dz_2\cdot \partial_2),
\end{equation}
and it is easy to check that 
\[
(V-V_{\text{div}})\biggl[\frac{\delta(a_1+a_2+a_3-1)}{\sum_{1\leq i<j\leq 3}a_ia_j(z_i-z_j)^2}\biggr]\xrightarrow{a_3\to 0} 0.
\]
Therefore, after trivially integrating $dz_i\cdot \partial_i = dt_i\partial_{t_i}$, the result of the integrand with the numerator $V_{\text{div}}$ is given by
\[
\int_{z_2,z_3} (dz_2\cdot dz_3)(T_\epsilon(x_2,z_2,z_3)-T_\epsilon(x_1,z_2,z_3))-\int_{z_1,z_3} (dz_3\cdot dz_1)(T_\epsilon(z_1,x_3,z_3)-T_\epsilon(z_1,x_2,z_3)),
\]
where only the first and fourth term diverge when $\epsilon\to 0^+$. From the definition of $M^\pm$ eq.\eqref{Mp} and eq.\eqref{Mm}, we see that these two divergent integrals are simply $M_{2j}^+$ and $M_{2j}^-$, so
\begin{equation}\label{divS1}
I_{\text{S}}(1,2,j)=M_{2j}^++M_{2j}^-+O(1).
\end{equation}
One can compute the $O(1)$ part from the generic star diagram in our cutoff regularization easily.

Similarly, for the star diagram with two cusps, we find that 
\begin{equation}\label{divS2}
\begin{aligned}
I_{\text{S}}(1,2,3)&=M_{23}^-+M_{31}^++\frac{\zeta_2}{4\epsilon^2}+\frac{\zeta_3}{\epsilon}+\frac{\zeta_2}{4\epsilon}\log(x_{13}^2x_{24}^2) +O(1)\\
&=M_{23}^-+M_{31}^++\frac{\zeta_2}{2}\log(\eta)^2-\frac{\zeta_2}{2}\log (\eta ) \log (x_{13}^2x_{24}^2)-3\zeta_3\log(\eta) +O(1).
\end{aligned}
\end{equation}
As we will explain later, the $\zeta_2$ and $\zeta_3$ contribution should be absorbed into the one-loop contribution in eq.\eqref{2Lremainder}, and the dependence of regularization of the $O(1)$ part should also be cancelled, which means at least the symbol of two $O(1)$ parts are the same.

\subsection{Y diagrams}\label{Ydiag}

Y diagram contains two points on one edge and one three-particle gluon vertex, 
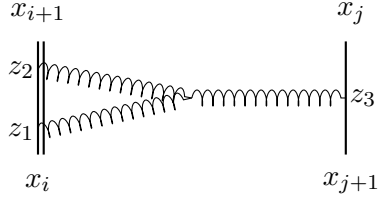
\begin{center}
\begin{tikzpicture}[scale=0.75]
\draw[thick] (-3,2.5) -- (-3,0.5);
\draw[thick] (2.4,2.5) -- (2.4,0.5);
\draw[thick] (-2.9,2.5) -- (-2.9,0.5);
\draw[decorate,decoration={aspect=0.3,coil,segment length=4,amplitude=3}] (-3,0.9) -- (-0.3,1.5);
\draw[decorate,decoration={aspect=0.3,coil,segment length=4,amplitude=3}] (-0.3,1.5) -- (2.4,1.5);
\draw[decorate,decoration={aspect=0.3,coil,segment length=4,amplitude=3}] (-3,2) -- (-0.3,1.5);
\node at (-3.3,0.9) {$z_1$};
\node at (-3.3,2) {$z_2$};
\node at (2.7,1.5) {$z_3$};
\node at (-3,0) {$x_{i}$};
\node at (-3,3) {$x_{i+1}$};
\node at (2.5,3) {$x_{j}$};
\node at (2.5,0) {$x_{j+1}$};
\end{tikzpicture}
\end{center}
so there is an ordering $z_1<z_2$. We first consider the full periodic version of the integrand
\[
\mathsf{I}_{\text{Y}}= \sum_{m,n\in\mathbb Z}\int_{z_1<z_2}\int d^4 x \,V\, T_\epsilon(z_1+mq,z_2+nq,z_3).
\]
Since $dz_1\cdot dz_2=p_1^2dt_1dt_2=0$, the numerator becomes
\[ 
V_{\text{Y}}=(dz_1\cdot dz_3) dz_2\cdot \partial_2-(dz_2\cdot dz_3) dz_1\cdot \partial_1, 
\]
which is exactly the $V_{\text{div}}$ defined in the last subsection. After integrating $dz_i\cdot \partial_i = dt_i\partial_{t_i}$ and renaming the remaining $z_2$ to $z_1$, we get that
\[
\begin{aligned}
\mathsf{I}_{\text{Y}}&=
\sum_{m,n\in\mathbb Z}\int_{x_2}^{y_2}\int_{x_1}^{y_1}(dz_1\cdot dz_3) (T_\epsilon(z_1+mq,y_1+nq,z_3)+T_\epsilon(x_1+mq,z_1+nq,z_3)\\
&\hspace{25ex}-2T_\epsilon(z_1+mq,z_1+nq,z_3))\\
&=\quad \begin{tikzpicture}[scale=0.5,baseline={([yshift=-0.5ex]current bounding box.center)}]
\draw[thick] (-3,2.6) -- (-3,0.4);
\draw[thick] (2.4,2.6) -- (2.4,0.4);
\draw[thick] (-3.1,2.6) -- (-3.1,0.4);
\draw (-3,1.4) -- (-0.3,1.5);
\draw (-0.3,1.5) -- (2.4,1.5);
\draw (-3,2.6) -- (-0.3,1.5);
\node[fill=black,circle,inner sep=2pt] at (-0.3,1.5) {};
\end{tikzpicture}
\,\, +\,\,
\begin{tikzpicture}[scale=0.5,baseline={([yshift=-0.5ex]current bounding box.center)}]
\draw[thick] (-3,2.6) -- (-3,0.4);
\draw[thick] (2.4,2.6) -- (2.4,0.4);
\draw[thick] (-3.1,2.6) -- (-3.1,0.4);
\draw (-3,0.4) -- (-0.3,1.5);
\draw (-0.3,1.5) -- (2.4,1.5);
\draw (-3,1.7) -- (-0.3,1.5);
\node[fill=black,circle,inner sep=2pt] at (-0.3,1.5) {};
\end{tikzpicture}
\,\, -\,\,
\cdots\\
&=\mathsf{M}_{i+1,j}^-+\mathsf{M}^+_{i,j}-2\sum_{m,n\in\mathbb Z}\int_{x_2}^{y_2}\int_{x_1}^{y_1}(dz_1\cdot dz_3)T_\epsilon(x_1+mq,z_1+nq,z_3)
\end{aligned}
\]

Note that the integral
\[
\int_{x_1}^{y_1}\int_{x_2}^{y_2}(dz_1\cdot dz_3) T_\epsilon(z_1+mq,z_1+nq,z_3)
\]
is a weight-3 polylogarithm that breaks the maximal transcendentality, and for $m=n$, it is further divergent. For $m=n$, as pointed out in \cite{Anastasiou:2009kna}, this weight-3 part is cancelled by the self-energy diagram (at $g^4$),
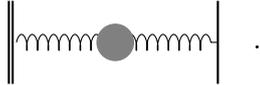
\[
~\quad \begin{tikzpicture}[scale=0.5,baseline={([yshift=-0.5ex]current bounding box.center)}]
\draw[thick] (-3,2.6) -- (-3,0.4);
\draw[thick] (2.4,2.6) -- (2.4,0.4);
\draw[thick] (-3.1,2.6) -- (-3.1,0.4);
\draw[decorate,decoration={aspect=0.3,coil,segment length=4,amplitude=3}] (-2.9,1.5) -- (-0.3,1.5);
\draw[decorate,decoration={aspect=0.3,coil,segment length=4,amplitude=3}] (-0.3,1.5) -- (2.4,1.5);
\node[fill=gray,circle,inner sep=5pt] at (-0.3,1.5) {};
\end{tikzpicture}\,\quad .
\]
A similar thing happens here in a periodic way, the propagators and vertices in the gray ball are periodic, which should cancel all $T_\epsilon(z_1+mq,z_1+nq,z_3)$, so we never consider these low-weight terms.

Again, we need to project the summation and only consider diagrams that can be contained in one period, which not only contains diagrams with $m=n$, 
\begin{equation}\label{divY1}
I_{\text{Y},1}(i,j)=M_{i+1,j}^-+M_{i,j}^+\,=\,\begin{tikzpicture}[scale=0.5,baseline={([yshift=-0.5ex]current bounding box.center)}]
\draw[thick] (-3,2.6) -- (-3,0.4);
\draw[thick] (2.4,2.6) -- (2.4,0.4);
\draw (-3,1.4) -- (-0.3,1.5);
\draw (-0.3,1.5) -- (2.4,1.5);
\draw (-3,2.6) -- (-0.3,1.5);
\node[fill=black,circle,inner sep=2pt] at (-0.3,1.5) {};
\end{tikzpicture}
\,\, +\,\,
\begin{tikzpicture}[scale=0.5,baseline={([yshift=-0.5ex]current bounding box.center)}]
\draw[thick] (-3,2.6) -- (-3,0.4);
\draw[thick] (2.4,2.6) -- (2.4,0.4);
\draw (-3,0.4) -- (-0.3,1.5);
\draw (-0.3,1.5) -- (2.4,1.5);
\draw (-3,1.7) -- (-0.3,1.5);
\node[fill=black,circle,inner sep=2pt] at (-0.3,1.5) {};
\end{tikzpicture}\quad ,
\end{equation}
but also contains some diagrams with $|m-n|=1$,
\begin{equation}
I_{\text{Y},2}(i,j)\,:=\,\begin{tikzpicture}[scale=0.5,baseline={([yshift=-0.5ex]current bounding box.center)}]
\draw[thick] (-2.9,2.5) -- (-2.9,0.5);
\draw[thick] (-1.5,3.5) -- (1,3.5);
\draw[thick] (2.4,2.5) -- (2.4,0.5);
\draw (-2.9,2.5) -- (-0.3,1.5);
\draw (-0.3,1.5) -- (2.4,1.5);
\draw (-0.3,3.5) -- (-0.3,1.5);
\node[fill=black,circle,inner sep=2pt] at (-0.3,1.5) {};
\end{tikzpicture}
\,\, +\,\,
\begin{tikzpicture}[scale=0.5,baseline={([yshift=-0.5ex]current bounding box.center)}]
\draw[thick] (-2.9,2.5) -- (-2.9,0.5);
\draw[thick] (-1.5,3.5) -- (1,3.5);
\draw[thick] (2.4,2.5) -- (2.4,0.5);
\draw (-2.9,1.5) -- (-0.3,1.5);
\draw (-0.3,1.5) -- (2.4,2.5);
\draw (-0.3,3.5) -- (-0.3,1.5);
\node[fill=black,circle,inner sep=2pt] at (-0.3,1.5) {};
\end{tikzpicture}\quad ,
\end{equation}
which are not divergent unless they contain a cusp, and the corresponding divergence reads
\begin{equation}\label{divY2}
I_{\text{Y},2}(i,i+1)=M^+_{i+1,i+n}+O(1),\quad 
I_{\text{Y},2}(i,i+n-1)=M^-_{i+n,i}+O(1).
\end{equation}

\subsection{X and XC diagrams}\label{Xdiag}

For X Diagrams
\[
I_{\text{X}}(i,j):=\begin{tikzpicture}[scale=0.75,baseline={([yshift=-0.5ex]current bounding box.center)}]
\draw[thick] (-2.5,3.5) -- (-3,1);
\draw[thick] (1.5,1) -- (1,3.5);
\draw[decorate,decoration={aspect=0.3,coil,segment length=4,amplitude=3}] 
        (-2.9,1.5) .. controls (-1.5,1) and (0,3) .. (1.1,3.2);
\node[fill=white,circle,inner sep=6.8pt] at (-0.8,2.1) {};
\draw[decorate,decoration={aspect=0.3,coil,segment length=4,amplitude=3}] 
        (1.4,1.4) .. controls (0,1) and (-2,3.5) .. (-2.6,3.1);
\node at (-3.3,1.5) {$z_1$};
\node at (-2.9,3.1) {$z_2$};
\node at (1.4,3.3) {$z_3$};
\node at (1.7,1.3) {$z_4$};
\end{tikzpicture}
\]
its integrand is
\begin{equation}
I_{\text{X}}= 4\int_{z_1<z_2}\int_{z_3<z_4}\frac{(dz_1\cdot dz_3)(dz_2\cdot dz_4)}{(z_1-z_3)^2(z_2-z_4)^2},
\end{equation}
and it is easy to integrate it to a weight-4 polylogrithm.

Besides, in the periodic summation of X diagrams, we have the following diagram where there is path ordering on points on different edges but in the same periodic class,
\[
\begin{tikzpicture}[baseline={([yshift=-3.5ex]current bounding box.center)},scale=0.75]
\draw[thick] (-2.9,2.5) -- (-2.9,0.5);
\draw[thick] (-1.5,3.5) -- (1,3.5);
\draw[thick] (-2.8,2.5) -- (-2.8,0.5);
\draw[decorate,decoration={aspect=0.3,coil,segment length=4,amplitude=3}] 
        (-2.9,1.8) .. controls (-1.5,1) and (0.5,1.5) .. (0.5,3.5);
\node[fill=white,circle,inner sep=7pt] at (-1,1.5) {};
\draw[decorate,decoration={aspect=0.3,coil,segment length=4,amplitude=3}] 
        (-2.8,1) .. controls (-1.2,0.7) and (-0.5,1.5) .. (-0.5,3.5);
\node at (-3.6,1.8) {$z_1(t_1)$};
\node at (-0.65,3.9) {$z_2(t_2)$};
\node at (0.65,3.9) {$z_3(t_3)$};
\node at (-3.6,1) {$z_4(t_4)$};
\end{tikzpicture}\qquad \longrightarrow\qquad
\begin{tikzpicture}[baseline={([yshift=-3.5ex]current bounding box.center)},scale=0.75]
\draw[thick] (-2.9,2.5) -- (-2.9,0.5);
\draw[thick] (-1.5,3.5) -- (1,3.5);
\draw[thick] (2.4,2.5) -- (2.4,0.5);
\draw[decorate,decoration={aspect=0.3,coil,segment length=4,amplitude=3}] 
        (2.4,2) .. controls (1,1) and (-0.5,1.5) .. (-0.5,3.5);
\node[fill=white,circle,inner sep=6.6pt] at (-0.1,2) {};
\draw[decorate,decoration={aspect=0.3,coil,segment length=4,amplitude=3}] 
        (-2.9,2) .. controls (-1.5,1) and (0.5,1.5) .. (0.5,3.5);
\node at (-3.6,2) {$z_1(t_1)$};
\node at (-0.65,3.9) {$z_2(t_2)$};
\node at (0.65,3.9) {$z_3(t_3)$};
\node at (3.2,1.9) {$z_4(t_4)$};
\node at (-3.1,0.3) {$2$};
\node at (-3.1,2.7) {$3$};
\node at (-1.8,3.5) {$5$};
\node at (1.4,3.5) {$6$};
\node at (2.6,2.7) {$8$};
\node at (2.6,0.3) {$9$};
\end{tikzpicture}
\]
which also looks like a curtain diagram, so we call it a XC diagram.
We define the dual points as 
\[
x_3-x_2=x_9-x_8=p_1, \quad x_6-x_5=p_2, \quad x_8-x_2=q,
\]
where $p_1$ and $p_2$ are massless, and then the integral is
\begin{equation}
I_{\text{XC}}(i,j)= \int_0^1dt_3\int_0^{t_3}dt_2\int_{0}^1dt_1\int_0^{t_1} dt_4\frac{((p_1+ p_2)^2)^2}{(z_1-z_3)^2(z_2-z_4)^2},
\end{equation}
where there is a constraint on $x_{i,j}^2$'s from $x_3-x_2=x_9-x_8=p_1$, so that
\[
-x_{2,5}^2+x_{2,6}^2+x_{3,5}^2-x_{3,6}^2=
-x_{5,8}^2+x_{5,9}^2+x_{6,8}^2-x_{6,9}^2=
(p_1+p_2)^2.
\]
Note that since $z_4<z_1+q$ from $t_4<t_1$, the diagram is seen as a diagram contained in one period and survives in the projection. The integration is straightforward, and its result has 5 independent algebraic letters of the form $X(a):=(z-a)/(\bar z-a)$ for
\[
a\in\biggl\{
0,\, 1,\, \frac{x_{5,8}^2}{x_{2,5}^2},\,\frac{ x_{6,8}^2}{x_{2,6}^2},\,\frac{x_{3,6}^2 x_{5,8}^2-x_{3,5}^2 x_{6,8}^2}{x_{2,5}^2 x_{3,6}^2-x_{2,6}^2 x_{3,5}^2}
\biggr\},
\]
and $z$, $\bar z$ are defined by
\begin{equation}\label{invuv}
z\bar z=\frac{x_{5,8}^2 x_{6,9}^2-x_{5,9}^2 x_{6,8}^2}{x_{2,5}^2 x_{3,6}^2-x_{2,6}^2 x_{3,5}^2},\quad (1-z)(1-\bar z)=\frac{4 (p_1\cdot q)(p_2\cdot q)}{x_{2,5}^2 x_{3,6}^2-x_{2,6}^2 x_{3,5}^2}
\end{equation}
In the case that
$x_{3,5}^2=0$ or $x_{6,8}^2=0$ (it only appears at $5$-pt), only 4 letters left, if further
$x_{3,5}^2=x_{6,8}^2=0$ (it only appears at $4$-pt), only first three letters left.

The algebraic part of its symbol with these algebraic letters is of the form
\begin{equation}\label{Fanti4m}
\frac{1}{2}\,F\otimes \biggl(\frac{1-z}{1-\bar z}\otimes (z\bar z)-\frac{z}{\bar z}\otimes ((1-z)(1-\bar z))\biggr),
\end{equation}
where $F$ is a weight-2 symbol,
\[
\begin{aligned}
F=\,&\frac{x_{5,9}^2 x_{6,8}^2}{x_{5,8}^2 x_{6,9}^2}\otimes X(0)-
\frac{x_{2,5}^2 x_{5,9}^2}{x_{3,5}^2 x_{5,8}^2}\otimes X\biggl(\frac{x_{5,8}^2}{x_{2,5}^2}\biggr)\\
&-\frac{x_{3,6}^2 x_{6,8}^2}{x_{2,6}^2 x_{6,9}^2}\otimes X\biggl(\frac{x_{6,8}^2}{x_{2,6}^2}\biggr)-
\frac{x_{3,5}^2 x_{6,9}^2}{x_{3,6}^2 x_{5,9}^2}\otimes X\biggl(\frac{x_{3,6}^2 x_{5,8}^2-x_{3,5}^2 x_{6,8}^2}{x_{2,5}^2 x_{3,6}^2-x_{2,6}^2 x_{3,5}^2}\biggr).
\end{aligned}
\]
It is interesting to see that the last two entries of eq.\eqref{Fanti4m} is the antipode of the symbol of the scalar one-loop triangle integral
\[
\mathcal S[(z-\bar z)T_0(z,\bar z)]=\frac{1}{2}\biggl((z\bar z)\otimes \frac{1-z}{1-\bar z}-((1-z)(1-\bar z))\otimes \frac{z}{\bar z}\biggr),
\]
which is also the well known four-mass box function.

The divergence of XC diagrams is also similar to curtain diagrams,
\begin{equation}\label{divXC}
I_{\text{XC}}(i,i+1)=2M^+_{i+1,i+n}+O(1),\quad I_{\text{XC}}(i,i+n-1)=2M^-_{i+n,i}+O(1).
\end{equation}
Note that for $I_{\text{XC}}(i,i+1)$, the regularization is introduced as $0<t_4<t_1<1-\delta_{i,i+2}$ for two points on different edges.


\section{Two-loop MHV form factors}

In this section, we give the $n$-particle two-loop integrand of the form factor from Wilson loop diagrams, make some comments, and then compute the two-loop 5- and 6-particle form factors as examples. 

\subsection{The integrand of remainder function and the cancellation of divergence}

Considering all possible diagrams, we write the two--loop remainder function as
\begin{align}\label{integrand}
R^{(2)}_n= &\sum_{1<j<k\leq n}(I_{\text{S}}(1,j,k)-I_{\text{C},1}(1,j,k)-I_{\text{C},2}(1,j,k)-I_{\text{C},3}(1,j,k))\nonumber\\
&+\sum_{j=2}^{n}(I_{\text{Y},2}(1,j)-\frac{1}{2}(I_{\text{XC}}(1,j)+I_{\text{C},2}(1,j,n+1)))\\
&+\sum_{j=3}^n(I_{\text{Y},1}(1,j)+I_{\text{Y},1}(j-2,n)-I_{\text{X}}(1,j))\nonumber\\
&-\sum_{\substack{1<j<k<l\\2<k\leq n, 1<l-j<n}} {\hspace{-5.5ex}}^{'}{\hspace{5.5ex}}I_{\text{O}}(1,k)I_{\text{O}}(j,l) \quad - \quad(\cdots)\quad \quad +\quad 
\text{cyclic}\nonumber,
\end{align}
the relative sign between diagrams with and without 3-point vertex is given by the coefficient of $C_FC_A$ in their color factor eq.\eqref{colorfactor}. Note that we combine $I_{\text{XC}}(1,j)+I_{\text{C},2}(1,j,n+1)$ because they share the same diagram, and we need to take their average by introducing a factor $1/2$. We also introduce a symbol $\sum'$ to avoid overcounting caused by periodicity. For example, for $n=5$,
\[
\begin{aligned}
&I_{\text{O}}(1,3)I_{\text{O}}(4,7)=I_{\text{O}}(1,3)I_{\text{O}}(-1,2) \xrightarrow{\text{cyclic}^{2}} I_{\text{O}}(1,4)I_{\text{O}}(3,5),\\
&I_{\text{O}}(1,5)I_{\text{O}}(3,7)=I_{\text{O}}(1,5)I_{\text{O}}(-2,2) \xrightarrow{\text{cyclic}^{3}} I_{\text{O}}(1,5)I_{\text{O}}(4,8),
\end{aligned}
\]
so there are only $12$ tuples of $\{j,k,l\}$, not all $14$ tuples restricted only by conditions $1<j<k<l$, $2<k\leq n$, and $1<l-j<n$. The dot part $(\cdots)$ is used to cancel the one-loop divergence.
We will check the cancellation of all divergence and then rewrite the integrand in finite integrals, see eq.\eqref{integrand2}.

In the summation eq.\eqref{integrand}, we do not consider the following two-loop cusp diagrams
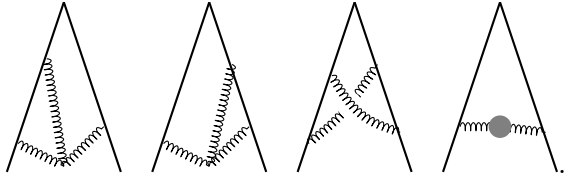
\[
\begin{tikzpicture}[scale=0.75]
\draw[thick] (-2,3.5) -- (-3,0.5);
\draw[thick] (-2,3.5) -- (-1,0.5);
\draw[decorate,decoration={aspect=0.3,coil,segment length=2,amplitude=1.5}] (-2.8,1) -- (-2,0.6);
\draw[decorate,decoration={aspect=0.3,coil,segment length=2,amplitude=1.5}] (-2,0.6) -- (-1.3,1.3);
\draw[decorate,decoration={aspect=0.3,coil,segment length=2,amplitude=1.5}] (-2.3,2.5) -- (-2,0.6);
\end{tikzpicture}
\quad 
\begin{tikzpicture}[scale=0.75]
\draw[thick] (-2,3.5) -- (-3,0.5);
\draw[thick] (-2,3.5) -- (-1,0.5);
\draw[decorate,decoration={aspect=0.3,coil,segment length=2,amplitude=1.5}] (-2.8,1) -- (-2,0.6);
\draw[decorate,decoration={aspect=0.3,coil,segment length=2,amplitude=1.5}] (-2,0.6) -- (-1.3,1.3);
\draw[decorate,decoration={aspect=0.3,coil,segment length=2,amplitude=1.5}] (-1.6,2.4) -- (-2,0.6);
\end{tikzpicture}
\quad 
\begin{tikzpicture}[scale=0.75]
\draw[thick] (-2,3.5) -- (-3,0.5);
\draw[thick] (-2,3.5) -- (-1,0.5);
\draw[decorate,decoration={aspect=0.3,coil,segment length=2,amplitude=1.5}] 
        (-2.8,1) .. controls (-2.5,1.3) and (-2.3,1.3) .. (-1.6,2.4);
\node[fill=white,circle,inner sep=3.4pt] at (-2.1,1.7) {};
\draw[decorate,decoration={aspect=0.3,coil,segment length=2,amplitude=1.5}] 
        (-2.4,2.2) .. controls (-2,1.5) and (-1.8,1.4) .. (-1.2,1.2);
\end{tikzpicture}
\quad 
\begin{tikzpicture}[scale=0.75]
\draw[thick] (-2,3.5) -- (-3,0.5);
\draw[thick] (-2,3.5) -- (-1,0.5);
\draw[decorate,decoration={aspect=0.3,coil,segment length=2,amplitude=1.5}] (-2.7,1.3) -- (-2,1.3);
\draw[decorate,decoration={aspect=0.3,coil,segment length=2,amplitude=1.5}] (-2,1.3) -- (-1.2,1.2);
\node[fill=gray,circle,inner sep=3pt] at (-2,1.3) {};
\end{tikzpicture}.
\]
They are already computed in dimensional regularization; see \textit{e.g.} \cite{Brandhuber:2010bj}
\[
\frac{1}{4 \epsilon^2}\left[\frac{\Gamma(1-2 \epsilon) \Gamma(1+\epsilon)}{\Gamma(1-\epsilon)}-1\right]\frac{(-s_{i,i+1})^{2 \epsilon}}{(2\epsilon)^2} ,
\]
so their sum is
\begin{equation}
\frac{1}{4\epsilon^2}\left[\frac{\Gamma(1-2 \epsilon) \Gamma(1+\epsilon)}{\Gamma(1-\epsilon)}-1\right]\sum_{i=1}^n\frac{(-s_{i,i+1})^{2 \epsilon}}{(2\epsilon)^2}\propto \zeta_2\mathcal{F}_n^{(1)}(2 \epsilon)+O(\epsilon),
\end{equation}
and it should be cancelled by $-f^{(2)}(\epsilon) \mathcal{F}_n^{(1)}(2 \epsilon)$ in the definition of the two-loop remainder function eq.\eqref{2Lremainder}. Besides, diagrams containing two adjacent cusps, $I_{\text{C},2}(1,2,3)$ and $I_{\text{S}}(1,2,3)$, have the $\zeta_3\epsilon^{-1}$ and $\zeta_2\epsilon^{-2}$ divergence together with some $\log(s_{i,i+1})$, see eq.\eqref{divC} and eq.\eqref{divS2}, and they should also be cancelled by the one-loop part. In fact, this cancellation is local on the Wilson loop side which involves at most two cusps, so it behaves the same as remainder functions of MHV amplitudes, see \cite{Drummond:2007bm} for an example. 
If one focuses on the symbol of the remainder function, these terms that are proportional to zeta values are already modded.
The nontrivial cancellation of divergence that depends on kinematic variables happens at the order $\epsilon^{-1}$ or $\log(\eta)$, it requires the sum of all $M^\pm$ contribution from divergent diagrams in eq.\eqref{integrand} to vanish.

To see it, we define the cyclic summation of  $M^\pm_{ij}$ with a fixed difference $j-i=k$ that
\begin{equation}\label{Mcyc}
\mathcal M_{k}^\pm := \sum_{i=1}^n M_{i,i+k}^\pm,
\end{equation}
where $k\in\mathbb Z$ can be negative, and all divergent contribution of $M^{\pm}$ from different diagrams is listed as follows.

The divergent part of all Y diagrams is given by eq.\eqref{divY1} and \eqref{divY2}
\[
\begin{aligned}
&\sum_{i=3}^{n}(I_{Y,1}(1,i)+I_{\text{Y},1}(i-2,n))+\text{cyclic}=
\sum_{k=2}^{n-1}(\mathcal M^+_k+\mathcal M_{-k}^++\mathcal M^-_{k-1}+\mathcal M_{-k-1}^-)+O(1),\\
&I_{\text{Y},2}(1,2)+I_{\text{Y},2}(1,n)+\text{cyclic}=\mathcal M_{n-1}^+ +\mathcal M^-_{-n}+O(1),
\end{aligned}
\]
the divergent part of divergent star diagrams is given by eq.\eqref{divS1} and  eq.\eqref{divS2} 
\[
\begin{aligned}
I_{\text{S}}(1,2,3)&+\sum_{i=4}^{n}I_{\text{S}}(1,2,i)+\sum_{i=1}^{n-3}I_{\text{S}}(i,n-1,n)+\text{cyclic}\\
&=\mathcal M_{1}^-+\mathcal M_{-2}^++\sum_{k=2}^{n-2}(\mathcal M_{k}^++\mathcal M_{-k-1}^++\mathcal M_{k}^-+\mathcal M_{-k-1}^-)+O(1),
\end{aligned}
\]
and from eq.\eqref{divC} three types of curtain diagrams give 
\[
\begin{aligned}
&\sum_{i=3}^{n}I_{\text{C},1}(1,2,i)+\sum_{i=4}^{n}I_{\text{C},2}(1,2,i)+\sum_{i=1}^{n-3}I_{\text{C},2}(i,n-1,n)+\sum_{i=1}^{n-2}I_{\text{C},3}(i,n-1,n))+\text{cyclic}\\
=&\,2\sum_{k=1}^{n-2}\mathcal M^-_k+2\sum_{k=2}^{n-2}\mathcal M^+_k+2\sum_{k=1-n}^{-3}\mathcal M^-_k+2\sum_{k=1-n}^{-2}\mathcal M^+_k+O(1).
\end{aligned}
\]
From eq.\eqref{divXC} and eq.\eqref{divC}, we know that XC diagrams have divergence
\[
\frac12(I_{\text{XC}}(1,2)+I_{\text{XC}}(1,n)+I_{\text{C},2}(1,2,n+1)+I_{\text{C},2}(1,n,n+1))+\text{cyclic}=2(\mathcal M_{n-1}^+ +\mathcal M^-_{-n})+O(1).
\]
Finally, it is easy to see that the integrand for the $n$-particle form factor is free of divergence by adding them together
\begin{equation}
\text{div}_{\text{Y},1}+\text{div}_{\text{Y},2}+\text{div}_{\text{S}}-\text{div}_{\text{C},1}-\text{div}_{\text{C},2}-\text{div}_{\text{C},3}-\text{div}_{\text{XC}}=0.
\end{equation}

Therefore, we can rewrite the integrand as
\begin{align}\label{integrand2}
R^{(2)}_n= &\sum_{1<j<k\leq n}(\hat I_{\text{S}}(1,j,k)-\hat I_{\text{C},1}(1,j,k)-\hat I_{\text{C},2}(1,j,k)-\hat I_{\text{C},3}(1,j,k))\nonumber\\
&+\sum_{j=2}^{n}(\hat I_{\text{Y},2}(1,j)-\frac{1}{2}(\hat I_{\text{XC}}(1,j)+\hat I_{\text{C},2}(1,j,n+1)))-\sum_{j=3}^n I_{\text{X}}(1,j)\\
&-\sum_{\substack{1<j<k<l\\2<k\leq n, 1<l-j<n}} {\hspace{-5.5ex}}^{'}{\hspace{5.5ex}}I_{\text{O}}(1,k)I_{\text{O}}(j,l)\quad +\quad 
\text{cyclic}\nonumber\nonumber,
\end{align}
where $\hat I_{*}$ is the regulated version of $I_{*}$ by subtracting the corresponding $M^\pm$, so all $\hat I_{\text{Y},1}=0$, and for diagrams with two adjacent cusps, we further subtract their one-loop divergence as
\[
\hat I_{\text{C},2}(i-1,i,i+1)=0,\quad \hat I_{\text{S}}(i-1,i,i+1)=O(1).
\]
We checked that this integrand at $n=4$ gives the known two-loop 4-particle remainder function! 

\subsection{Some results of two-loop MHV form factors}

In this subsection, we list some results of the two-loop form factors obtained by the integrand given by the last section.

\paragraph{Square roots} We first count the number of square roots. For $R_n^{(2)}$, one-loop triangle square roots come from star diagrams, and after summing all star diagrams, we find that one-loop square roots only left the one appearing in one-loop triangle Feynman diagrams
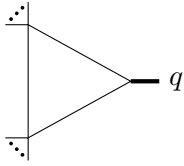
\begin{center}
\begin{tikzpicture}[baseline={([yshift=-.5ex]current bounding box.center)},scale=0.75]
\draw[ultra thick] (2.8,0.2) -- (3.3,0.2);
\node[inner sep= .5pt, fill = black, circle] at (0.7,1.3) {};
\node[inner sep= .5pt, fill = black, circle] at (0.7,-0.9) {};
\node[inner sep= .5pt, fill = black, circle] at (0.8,-1) {};
\node[inner sep= .5pt, fill = black, circle] at (0.9,-1.1) {};
\node[inner sep= .5pt, fill = black, circle] at (0.9,1.5) {};
\node[inner sep= .5pt, fill = black, circle] at (0.8,1.4) {};
\draw (1,1.2) -- (1,-0.8) -- (2.8,0.2) -- cycle;
\draw (0.6,-0.8) -- (1,-0.8) -- (1,-1.2);
\draw (0.6,1.2) -- (1,1.2) -- (1,1.6);
\node at (3.6,0.2) {$q$};
\end{tikzpicture}
\end{center}
that is defined by equations
\[
z\bar z=\frac{(p_i+\cdots+p_j)^2}{q^2},\quad 
(1-z)(1-\bar z)=\frac{(p_{j+1}+\cdots+p_{i-1})^2}{q^2}.
\]
All other one-loop triangle square roots are cancelled in the summation, which also explains why these square roots do not appear in the two-loop MHV amplitudes. 
Therefore, there are $n(n-3)/2$ one-loop square roots for $R_n^{(2)}$, 
where the factor $1/2$ comes from the flip symmetry. 

All two-loop square roots come from the XC diagram
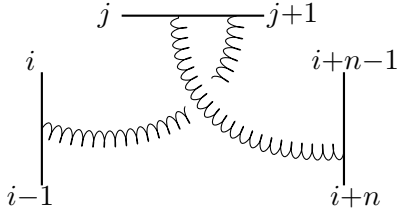
\[
\begin{tikzpicture}[baseline={([yshift=-3.5ex]current bounding box.center)},scale=0.75]
\draw[thick] (-2.9,2.5) -- (-2.9,0.5);
\draw[thick] (-1.5,3.5) -- (1,3.5);
\draw[thick] (2.4,2.5) -- (2.4,0.5);
\draw[decorate,decoration={aspect=0.3,coil,segment length=4,amplitude=3}] 
        (-2.9,1.5) .. controls (-1.5,1) and (0.5,1.5) .. (0.5,3.5);
\node[fill=white,circle,inner sep=6pt] at (0,2) {};
\draw[decorate,decoration={aspect=0.3,coil,segment length=4,amplitude=3}] 
        (2.4,1.1) .. controls (1,1) and (-0.5,1.5) .. (-0.5,3.5);
\node at (-3.1,0.3) {$i{-}1$};
\node at (-3.1,2.7) {$i$};
\node at (-1.8,3.5) {$j$};
\node at (1.5,3.5) {$j{+}1$};
\node at (2.6,2.7) {$i{+}n{-}1$};
\node at (2.6,0.3) {$i{+}n$};
\end{tikzpicture}
\]
it requires that $i<j<i+n-2$, so there are also $n(n-3)/2$ two-loop square roots for the $n$-particle form factor, where we also need to divide $2$ according to the flip symmetry.

\paragraph{$R_5^{(2)}$ and $R_6^{(2)}$} Now we come to two instances, the two-loop 5-particle and 6-particle form factor. We computed their symbol, which can be found in the ancillary  files. 

To check their correctness, we apply some collinear limits.
The first non-trivial check is that as two external particles become collinear $n|\!|1$, the $n$-particle form factor becomes a $(n-1)$-particle form factor times the splitting function that is captured by the BDS ansatz in planar $\mathcal N=4$ sYM, so the remainder function behaves trivially as $R_n\to R_{n-1}$. We checked that our result of the two-loop remainder function of 6/5-particle form factor becomes the 5/4-particle one, which matches the known result of the 4-particle form factor~\cite{Dixon:2022xqh,Dixon:2024yvq,Guo:2024bsd}. 

If we further take more than two adjacent legs to be collinear, the form factor factorizes to the product of two different objects. To take the limit, it is more convenient to use the OPE variables~\cite{Sever:2020jjx,Sever:2021nsq,Sever:2021xga}, in which different collinear limits are sending some $T_i\to 0$, 
we list the OPE variables up to 6-particle in Appendix \ref{OPE}. For the 5-particle form factor, the triple collinear limit $n|\!|1|\!|2$ ($T_2\to 0$) becomes $(R_3\times A_6)^{(2)}=R_3^{(2)}+A_6^{(2)}$ (because the one-loop remainder function is trivial) and the quadruple collinear limit $n|\!|1|\!|2|\!|3$ ($T_1\to 0$) becomes $(R_2\times A_7)^{(2)}=A_7^{(2)}$, where $A_n$ is the reminder function of the $n$-particle MHV amplitude.
We checked both limits and find that they match the known result of the two-loop form factors and the MHV amplitudes~\cite{Dixon:2022rse,Caron-Huot:2011zgw}.

For the two-loop $5$-particle form factor, there are $135$ letters in its symbol. There are $5$ one-loop and $5$ two-loop square roots, so $10$ square roots in total. In $135$ letters, there are $30$ letters with one-loop square roots and $20$ letters with two-loop square roots. In the remaining $85$ letters, there are $60$ parity-even rational letters, and the other $25$ rational letters are parity-odd. The parity of a letter $x$ here is its behavior under swapping  helicity spinors $\lambda_i$ and $\tilde\lambda_i$, it is parity-odd\footnote{In terms of planar variables $\{s_{i,\dots,j}\}$, parity-odd letters contain square roots of Gram determinant of external momenta.} if $x\to x^{-1}$ and it is parity-even if $x\to x$. The number of independent letters at each slot of the symbol is 15, 70, 105, 24. 
The similar counting for the 6-particle form factor can be read in the ancillary files, so we do not list it here.

\paragraph{Last entry condition} A more interesting observation is the last entry condition for MHV form factors. We find that the parity-even letters that can appear in the last entry are exactly cross-ratios on the right-hand side of eq.\eqref{invuv} 
\begin{equation}\label{les}
\biggl\{
\frac{x_{j,i^+}^2 x_{j+1,i^++1}^2-x_{j,i^++1}^2 x_{j+1,i^+}^2}{x_{i,j}^2 x_{i+1,j+1}^2-x_{i,j+1}^2 x_{i+1,j}^2},\quad \frac{4 (p_i\cdot q)(p_j\cdot q)}{x_{i,j}^2 x_{i+1,j+1}^2-x_{i,j+1}^2 x_{i+1,j}^2}
\biggr\}_{i,j},
\end{equation}
where $i^+:=i+n$. Since parity-odd letters only appear in the last two entries, these letters are exactly the last entries of the parity-even part of the remainder functions due to the Galois symmetry.
To see this condition in a more familiar form~\cite{CaronHuot:2011kk}, we rewrite it in momentum twistors; see Appendix \ref{momtwitor} for an very brief introduction to momentum twistors for a periodic Wilson loop. According to $x_{ij}^2\propto \langle i-1,i,j-1,j\rangle$, the two factors in the first letter read 
\[
x_{i,j}^2 x_{i+1,j+1}^2-x_{i,j+1}^2 x_{i+1,j}^2\propto \langle i-1,i,i+1,j\rangle \langle i,j-1,j,j+1\rangle,
\]
and 
\[
x_{j,i^+}^2 x_{j+1,i^++1}^2-x_{j,i^++1}^2 x_{j+1,i^+}^2\propto \langle j-1,j,j+1,i^+\rangle \langle j,i^+-1,i^+,i^++1\rangle,
\]
and from the periodicity of the spinor $\lambda_{i^+}^{\alpha}=\lambda_i^{\alpha}$ and eq.\eqref{ibarj}, we also have that
\[
2p_i\cdot q \propto \langle i-1,i,i+1,i^+\rangle.
\]
Therefore, we conclude that only $\langle i-1,i,i+1,j\rangle$ for some $i,j\in\mathbb Z$ can appear in the last entries of the parity-even part of the two-loop MHV form factors, which is exactly the same last entry condition for the MHV amplitudes. 

For parity-odd letters in last entries, we find that they are
\begin{equation}
\biggl\{\frac{\langle i,i+1\rangle}{[i,i+1]},\quad \frac{\sum_{k=i}^j\langle ik\rangle [kj]}{\sum_{k=i}^j [ik]\langle kj\rangle}\biggr\}_{i,j}
\end{equation}
for $1<j-i<n$. After taking out factors $\langle i-1,i,i+1,j\rangle$ and $x_{i,i+1}^2$ according to eq.\eqref{xmom} and eq.\eqref{ibarj}, the remaining factors can only be $\langle i,i+1\rangle$ for any $i$.

\paragraph{2D limit} Another interesting limit is the 2D limit, which is first considered at strong coupling both for amplitudes \cite{Alday:2009yn} and form factors \cite{Maldacena:2010kp}, and then at weaking coulping for amplitudes \cite{CaronHuot:2011kk,Caron-Huot:2013vda} and form factors \cite{Dixon:2024yvq}. The limit restricts external momentum living in a two-dimensional subspace (AdS$_{3}$ on the string side). Since 2D massless kinematics only has two directions $(1,1)$ and $(1,-1)$, which makes the Wilson loop a zigzag, so only the Wilson loop with even cusps can be extended periodically, which means all $(2n+1)$-particle 2D form factors are trivial. Therefore, we consider the 2D limit for the 6-particle form factor.

For the 2D kinematics, we parameterize helicity spinors as \cite{Elvang:2014fja}
\[
\tilde\lambda_i =\lambda_i =\sqrt{2e_i}\,\biggl(\frac{t_i^2-1}{t_i^2+1},\frac{2 t_i}{t_i^2+1}\biggr)^{\mathsf{T}},
\]
where 
\[
t_i = \begin{cases}
\alpha_i\epsilon, &\text{$i$ odd};\\
1-\alpha_i\epsilon, &\text{$i$ even},
\end{cases}
\]
for a small positive parameter $\epsilon\sim 0^+$, 
where $\{\alpha_i\}$ are random positive variables, and the $\epsilon\to 0^+$ limit should do not depend on them. The parameter $e_i$ is the energy of the $i$-th 2D massless particle. 

We take the 2D limit of the two-loop remainder function of 6-particle form factor, and find that its alphabet contains $21$ letters
\[
\begin{aligned}
\{&E_1,E_2,E_3,E_4,E_5,E_6,\\
&E_1+E_3,E_2+E_4,E_1+E_5,E_3+E_5,E_2+E_6,E_4+E_6,\\
&E_1-E_2,E_2-E_3,E_3-E_4,E_4-E_5,E_5-E_6,E_6-E_1,\\
&E_3-E_4+E_5,E_4-E_5+E_6,E_5-E_6+E_1\},
\end{aligned}
\]
where 
\[
E_{2i}=\frac{e_{2i}}{e_2+e_4+e_6},\quad E_{2i+1}=\frac{e_{2i+1}}{e_1+e_3+e_5}.
\]
The last $9$ letters only appear at the third slot in its symbol, and they come from letters with square roots in 4D. Therefore, unlike the 2D two-loop $n$-particle MHV amplitude \cite{CaronHuot:2011kk,Caron-Huot:2013vda} is a sum of the product of 4 logarithms, the form factor has other types of polylogarithms caused by letters with square roots. One can further get the 2D two-loop 4-particle form factor by the double soft limit $e_5,e_6\to 0$, whose function result is already given in Appendix B of \cite{Dixon:2024yvq}. 

Since there is no new type of square root for $n$-particle, we conjecture that letters of the 2D $n$-particle form factor are in the form
\begin{equation}\label{2Dletters}
\sum_{j=1}^kE_{i+2j},\quad \sum_{j=0}^k (-1)^jE_{i+j}
\end{equation}
for $i=1,\dots,n$ and $k=1,\dots,(n/2-1)$, where
\[
E_{2i}=\frac{e_{2i}}{\sum_{k=1}^{n/2} e_{2k}},\quad E_{2i+1}=\frac{e_{2i+1}}{\sum_{k=1}^{n/2} e_{2k-1}},
\]
with $\sum_i E_{2i}=\sum_i E_{2i+1}=1$. Similarly, we further conjecture that the second type of letters with minus signs in eq.\eqref{2Dletters} only appear at the third slot of the symbol. For the $2n$-particle case, there are $2n(n-1)$ letters of the first type, and $n(2n-3)$ letters of the second type, so there are $n(4n-5)$ letters in total.

\section{Conclusions}

In this note, we discussed how to perturbatively expand the periodic Wilson loop which is dual to the MHV form factor beyond the one-loop level.  Besides the ordinary diagrams, one also needs to consider diagrams with the path ordering of points in different edges but in the same periodic image, which corresponds to non-planar Feynman diagrams in the integrand of form factors. Following this rule, we determined the $n$-particle two-loop integrand eq.\eqref{integrand2}. For each two-loop diagram in the integrand, we showed how to regulate it and proved the cancellation of all divergence.
In particular, it is successful to regain the symbol of the known 4-particle result \cite{Dixon:2022xqh,Dixon:2024yvq,Guo:2024bsd}. We also computed two new data points, 5- and 6-particle form factors, and checked that they behave correctly under some collinear limits. After taking the 2D limit of the 6-particle form factors, we made a prediction of the symbol alphabet for the 2D two-loop $n$-particle form factor.


The $n$-particle two-loop integrand eq.\eqref{integrand2} sheds some light on the application of form factor/Wilson loop duality at weak coupling. One direct generalization is to further compute the higher-loop correction by the same rule as we stated, but a more interesting one is to consider the periodic \textit{super} Wilson loop, not only for the form factors of other $\frac{1}2$-BPS operators~\cite{Basso:2023bwv, Basso:2024hlx,Guo:2021bym,Henn:2024pki}, but also supersymmetry and dual supersymmetry may help to compute the MHV form factor in a more efficient way. As shown in \cite{Caron-Huot:2011zgw,CaronHuot:2011kk}, the anomaly of half of the dual super conformal symmetry helps to compute the two-loop MHV amplitudes from the one-loop NMHV amplitudes in a more symmetric way, where all one-loop triangle square roots in star diagrams do not appear in intermediate steps by the dual conformal symmetry. In addition, the anomaly method also restricts the last entry of all-loop MHV amplitudes. We also find a similar last entry condition on the two-loop MHV form factor, so we expect that a similar argument from the periodic super Wilson loop would prove this last entry condition at any loop.

Another interesting question is to understand the antipodal duality from the periodic (super) Wilson loop side. In \cite{Dixon:2022xqh}, it is found that the symbol of $R_4^{(2)}$ and $R_4^{(3)}$ becomes to itself by applying a kinematics map with the antipode of its symbol, and it is believed that the duality is an all-loop duality and holds beyond the symbol. In the XC diagram, we found a similar antipodal structure in its symbol. Its algebraic part eq.\eqref{Fanti4m} have an antipodal one-loop triangle (or four-mass box function) as its last two entries. Although it does not correspond to the known antipodal map, it seems that there may be some hidden antipodal structure in the symbol of some non-planar Feynman diagrams. Still focusing on the two-loop form factors, since we have the 5- and 6-particle form factor now, it is also interesting to look for the antipodal duality for form factors with higher multiplicities in the future.

\section*{Acknowledgement} The author thanks Lance Dixon, Song He, Yichao Tang, Shuo Xin, Qinglin Yang for useful discussions. The author is grateful to the Simons Center for Geometry and Physics for hospitality during the program “Solving $\mathcal N = 4$ super Yang-Mills theory via Scattering Amplitudes”. 
This work is supported by the U.S. Department of Energy under contract number DE-AC02-76SF00515.

\clearpage
\appendix

\section*{Appendix}

\section{Momentum twistor}\label{momtwitor}

We list some basic definitions of bosonic momentum twistors here. For a complete definition of momentum supertwistors, see \textit{e.g.} \cite{Basso:2023bwv}.

For a (finite or countably infinite) set of dual momenta $\{x_i\}$, two adjacent dual momenta are related by a massless momentum
\[
x_{i+1}-x_i=p_i=\lambda_i^{\alpha}\tilde\lambda_i^{\dot\alpha},
\]
and we define the momentum twistor as a projective 4-vector
\begin{equation}\label{eq:mt}
Z_i=\left(\begin{array}{c}
\lambda_i^\alpha \\
\mu_i^{\dot\alpha}
\end{array}\right)=\left(\begin{array}{c}
\lambda_i^\alpha \\
x_i^{\alpha \dot{\alpha}} \lambda_{i \alpha} 
\end{array}\right),
\end{equation}
which linearizes a conformal transformation of dual momenta $\{x_i\}$ to a $\operatorname{SL}(4)$ matrix acting on momentum twistors $\{Z_i\}$. The basic dual conformal $\operatorname{SL}(4)$ invariant is a determinant of four 4-vectors called Pl\"ucker coordinate
\begin{equation}
\langle i,j,k,l\rangle :=\epsilon_{ABCD}Z_i^AZ_j^BZ_k^CZ_l^D = \det([Z_i\, Z_j\, Z_k\, Z_l]).
\end{equation}
Pl\"ucker coordinates are not independent, they satisfy quadratic relations
\begin{equation}
\sum_{k=1}^5(-1)^k\langle i_1,i_2,i_3,j_k\rangle\langle j_1,\dots,\widehat{j_k},\dots,j_5\rangle = 0
\end{equation}
for any tuples $\{i_1,i_2,i_3\}$ and $\{j_1,j_2,j_3,j_4,j_5\}$.

The inverse map of eq.\eqref{eq:mt} is 
\begin{equation}
x_i^{\alpha \dot{\alpha}}=\frac{\lambda_{i-1}^\alpha \mu_i^{\dot{\alpha}}-\lambda_i^\alpha \mu_{i-1}^{\dot{\alpha}}}{\langle i-1,i\rangle}.
\end{equation}
From it, planar variables are related to Pl\"ucker coordinates by
\begin{equation}\label{xmom}
x_{i,j}^2=(x_i-x_j)^2=\frac{\langle i-1,i,j-1,j\rangle}{\langle i-1,i\rangle \langle j-1,j\rangle},
\end{equation}
where $\langle i,j\rangle :=\lambda_i^\alpha\lambda_j^\beta \epsilon_{\alpha\beta}$ is the angle bracket of two helicity spinors. The following identity is used in the note
\begin{equation}\label{ibarj}
\frac{\langle i-1,i,i+1,j\rangle}{\langle i-1,i\rangle\langle i,i+1\rangle} = \lambda_j^\alpha(x_j-x_i)_{\alpha\dot\alpha}\tilde \lambda_i^{\dot\alpha}.
\end{equation}
We can further introduce a bi-twistor 
\[
I_\infty = \begin{pmatrix}
 0 & 0 & 1 & 0 \\
 0 & 0 & 0 & 1 \\
\end{pmatrix}^{\mathsf{T}}
\]
to rewrite the angle bracket of two helicity spinors $\langle i,j\rangle$ to a 4-bracket $\langle i,j,I_\infty\rangle$. 

For the periodic Wilson loop dual to a $n$-particle form factor, $p_{i+n}=p_i$, so $\lambda_{i+n}=\lambda_{i}$ and $\tilde\lambda_{i+n}=\tilde\lambda_{i}$. However, $Z_{i+n}$ is not equal to $Z_i$ since $x_{i+n}=x_i+q$, and they are related by a universal linear map $\mathsf P\in \operatorname{SL}(4)$ as 
\begin{equation}
Z_{i+n}=\left(\begin{array}{c}
\lambda_{i+n}^\alpha \\
x_{i+n}^{\alpha \dot{\alpha}} \lambda_{(i+n)\, \alpha} 
\end{array}\right)=\left(\begin{array}{c}
\lambda_i^\alpha \\
(x_{i}+q)^{\alpha \dot{\alpha}} \lambda_{i \alpha} 
\end{array}\right)= \begin{pmatrix}
\delta^{\alpha}_\beta&0\\
q^{\phantom{\beta}\dot\alpha}_{\beta}&\delta^{\dot\alpha}_{\dot\beta}\\
\end{pmatrix}\left(\begin{array}{c}
\lambda_i^\beta \\
x_{i}^{\alpha \dot{\beta}} \lambda_{i \alpha} 
\end{array}\right)=\mathsf{P}Z_i.
\end{equation}
Note that the bi-twistor $I_{\infty}$ is invariant under the transition map
\[
\mathsf{P}I_{\infty}=I_{\infty},
\]
so this gives two independent eigenvectors with eigenvalue $1$, and there is no other non-zero independent eigenvector for a massive $q$.

It is allowed to define momentum twistors $\{Z_i\}$ up to a global $\operatorname{SL}(4)$ transformation because we only consider functions in Pl\"ucker coordinates $\langle i,j,k,l\rangle$. However, the transition map should be transformed as
\[
\mathsf{P}\mapsto \Lambda\mathsf{P}\Lambda^{-1} \quad \text{when}\quad Z_i \mapsto \Lambda Z_i,
\]
for a $\Lambda\in \operatorname{SL}(4)$.

If the period $q$ is further light-like, then $\{x_1,\dots,x_{n},x_{1^+}\}$ forms a closed light-like polygon, so we can assign a twistor $Z_{n+1}$ to the negative period $-q$, and the transition map becomes
\begin{equation}
\mathsf P(Z)=Z+\frac{\langle n+1,Z \rangle}{\langle n,n+1\rangle\langle n+1,1\rangle} I_\infty\cap (n,n+1,1),
\end{equation}
where 
\[
I_\infty\cap (n,n+1,1):=-\langle n,n+1\rangle Z_1+\langle n,1\rangle Z_{n+1}-\langle n+1,1\rangle Z_n.
\]

\section{OPE variables}\label{OPE}

The OPE variables are first introduced for the OPE of the form factors \cite{Sever:2020jjx,Sever:2021nsq,Sever:2021xga}, which is compatible for the near-collinear region $T_i\sim 0$. 

We choose the infinity bi-twistor and the transition matrix as
\[
I_\infty = \begin{pmatrix}
 1 & -1 & 0 & 0 \\
 0 & 0 & 1 & -1 \\
\end{pmatrix}^{\mathsf{T}},\quad \mathsf{P}=\begin{pmatrix}
 2 & 1 & 0 & 0 \\
 -1 & 0 & 0 & 0 \\
 0 & 0 & 2 & 1 \\
 0 & 0 & -1 & 0 \\
\end{pmatrix}.
\]
Note that the bi-twistor $I_{\infty}$ is choosed to be invariant under the transition map $\mathsf{P}I_{\infty}=I_{\infty}$.

We first define the following matrix
\[
M_1=\left(
\begin{array}{cccc}
 T_1 & 0 & 0 & 0 \\
 0 & T_1^{-1} & 0 & 0 \\
 0 & 0 & S_1 & 0 \\
 0 & 0 & 0 & S_1^{-1} \\
\end{array}
\right),\,\,
M_2=\frac1{\sqrt{F_2}S_2T_2}\left(
\begin{array}{cccc}
 F_2 S_2 & 0 & 0 & 0 \\
 -F_2 S_2 \left(T_2^2-1\right) & F_2 S_2 T_2^2 & 0 & S_2 T_2 \left(S_2-F_2 T_2\right) \\
 T_2-F_2 S_2 & 0 & T_2 & 0 \\
 0 & 0 & 0 & S_2^2 T_2 \\
\end{array}
\right),
\]
\[
M_3=\frac1{\sqrt{F_3}S_3T_3}
{\footnotesize
\begin{pmatrix}
 -S_3 (F_3 (1-3 T_3^2)+S_3 T_3) & -F_3 S_3 (T_3^2-1) & S_3 T_3 (F_3 T_3-S_3) & F_3 S_3 (T_3^2-1) \\
 3 F_3 S_3 (T_3^2-1)-(S_3^2-1) T_3 & -F_3 S_3 (T_3^2-3)-T_3 & S_3 T_3 (F_3 T_3-S_3) & F_3 S_3 (T_3^2-3)+2 T_3 \\
 S_3 (F_3 (1-3 T_3^2)+2 S_3 T_3) & F_3 S_3 (T_3^2-1) & S_3 T_3 (2 S_3-F_3 T_3) & -F_3 S_3 (T_3^2-1) \\
 T_3-F_3 S_3 & F_3 S_3-T_3 & 0 & 2 T_3-F_3 S_3 \\
\end{pmatrix}
},
\]
\[
\begin{aligned}
&M_4=\frac1{\sqrt{F_4}S_4T_4}\times\quad\\
&
{\scriptsize
\begin{pmatrix}
 F_4 S_4 \left(9-5 T_4^2\right)+\left(S_4^2-4\right) T_4 & 3 F_4 S_4 \left(T_4^2-1\right)-\left(S_4^2-1\right) T_4 & -F_4 S_4 \left(T_4^2-3\right)-2 T_4 & F_4 S_4 \left(3-4 T_4^2\right)+\left(2 S_4^2-1\right) T_4 \\
 4 \left(S_4^2-1\right) T_4-15 F_4 S_4 \left(T_4^2-1\right) & F_4 S_4 \left(9 T_4^2-5\right)+\left(1-4 S_4^2\right) T_4 & F_4 S_4 \left(5-3 T_4^2\right)-2 T_4 & F_4 S_4 \left(5-12 T_4^2\right)+\left(8 S_4^2-1\right) T_4 \\
 F_4 S_4 \left(5 T_4^2-12\right)-\left(S_4^2-8\right) T_4 & F_4 S_4 \left(4-3 T_4^2\right)+\left(S_4^2-2\right) T_4 & F_4 S_4 \left(T_4^2-4\right)+4 T_4 & 4 F_4 S_4 \left(T_4^2-1\right)-2 \left(S_4^2-1\right) T_4 \\
 S_4 \left(F_4 \left(3-5 T_4^2\right)+2 S_4 T_4\right) & -S_4 \left(F_4 \left(1-3 T_4^2\right)+2 S_4 T_4\right) & -F_4 S_4 \left(T_4^2-1\right) & S_4 \left(F_4 \left(1-4 T_4^2\right)+4 S_4 T_4\right) \\
\end{pmatrix}
}.
\end{aligned}
\]
For the 4-particle form factor, we define its momentum twistors as
\begin{equation}\label{OPE4}
\begin{aligned}
Z_1&=M_1 (0,1,0,1)^{\mathsf{T}},\quad 
Z_2=M_1 M_2 (1,3,-1,1)^{\mathsf{T}},\\
Z_3&=M_1 M_2 (1,1,-2,0)^{\mathsf{T}},\quad
Z_4=(0,0,1,0)^{\mathsf{T}}.
\end{aligned}
\end{equation}
For the 5-particle form factor, we define its momentum twistors as
\begin{equation}\label{OPE5}
\begin{aligned}
Z_1&=M_1 (0,1,0,1)^{\mathsf{T}},\quad 
Z_2=M_1 M_2 M_3 (1,4,-1,2)^{\mathsf{T}},\\
Z_3&=M_1 M_2 M_3 (3,5,-4,1)^{\mathsf{T}},\quad
Z_4=M_1 M_2 (1,1,-2,0)^{\mathsf{T}},\quad
Z_5=(0,0,1,0)^{\mathsf{T}}.
\end{aligned}
\end{equation}
For the 6-particle form factor, we define its momentum twistors as
\begin{equation}\label{OPE6}
\begin{aligned}
    Z_1&=M_1  (0,1,0,1)^{\mathsf{T}},\quad 
    Z_2=M_1 M_2 M_3 (1,4,-1,2)^{\mathsf{T}},\quad 
    Z_3=M_1 M_2 M_3 M_4 (5,12,-6,4)^{\mathsf{T}},\\
    Z_4&=M_1 M_2 M_3 M_4 (4,6,-6,1)^{\mathsf{T}},\quad 
    Z_5=M_1 M_2 (1,1,-2,0)^{\mathsf{T}},\quad 
    Z_6=(0,0,1,0)^{\mathsf{T}}.
\end{aligned}
\end{equation}

\bibliographystyle{utphys}
\bibliography{refs}

\end{document}